\documentclass[11pt,a4paper]{article}

\usepackage[a4paper,margin=1in]{geometry}
\usepackage{authblk}
\usepackage{setspace}

\usepackage{amsmath,amssymb}
\usepackage{graphicx,xcolor}
\usepackage{hyperref}
\usepackage{enumitem}
\setlist{nosep}

\usepackage{graphicx,graphics,subfig}
\usepackage{comment}

\usepackage{enumitem}
\setlist[enumerate]{leftmargin=0pt, labelwidth=!, labelsep=0.5em}

\usepackage{cleveref}

\graphicspath{{./figures/}}
\crefname{equation}{}{}  
\Crefname{equation}{Eq.}{Eqs.} 
\crefformat{equation}{(#2#1#3)} 
\Crefformat{equation}{Eq.~(#2#1#3)}

\title{From Phenomenology to a Nonlinear Model of Dynamic Snap-Through
  of an Elastica}

\author{Chiraprabha Bhattacharyya}
\author{Ramsharan Rangarajan\thanks{Corresponding author: \texttt{rram@iisc.ac.in}}}
\affil{Mechanical Engineering, Indian Institute of Science
  Bangalore, India}

\date{} 

\begin{document}
\maketitle



\begin{abstract}
  Rotating the clamped ends of a buckled elastica induces a
  snap-through instability. Predicting the limit point and determining
  the equilibria at the start and end of the snap are routine
  computations in the quasi-static setting. The instability itself,
  however, is dynamic, and quite violently so. We propose an
  energy-preserving nonlinear single degree of freedom model for this
  dynamic phenomenon in the case of a symmetrically deforming
  elastica. The model hinges on a surprising observation relating
  elastica profiles during the free dynamic snap with a specific
  sequence of geometrically-constrained elastic energy minimizing
  configurations. We corroborate this phenomenological observation
  over a significant range of arch depths through experiments and
  finite element simulations.  The resulting model does not rely on
  modal expansions, explicit slowness assumptions, or linearization of
  the arch's kinematics. Instead, the model is effective because its
  solutions approximate the action integral well. The model provides
  distinctive computational benefits and new insights on the
  snap-through phenomenon.  Our study is motivated by an application
  harnessing snap-through instabilities in submerged ribbons for
  underwater propulsion. We briefly describe its novel working
  principle and discuss its relationship to the problem studied.
\end{abstract}


\section{Introduction}
\label{sec:1}
Snap-through instabilities are a ubiquitous theme in the study of
slender structures.  A quintessential problem in this context is the
snapping of a curved arch subject to transverse loading
\cite{Pippard1990,Patrcio1998}. Typically, increasing the load drives
the structure to a limit point. With no equilibrium solution available
in the vicinity, the structure abruptly jumps to a non-adjacent
configuration. Such a discontinuous dependence of the solution on the
forcing is unlike the response seen in buckling instabilities, where
the structure gradually transitions to a new (symmetry-breaking)
solution branch past the bifurcation point.  While classical studies
primarily focused on determining critical loads to avoid snap-through,
the recent literature is replete with examples embracing them to
gainfully harness the rapid energy release possible
\cite{Hu2015,Reis2015}. Studies to this effect include, for instance,
devising MEMS switches \cite{Das2009, Ouakad2014}, designing soft
robots \cite{Pal2021, Tang2020, Wang2023}, energy harvesting
applications \cite{Nan2021}, reconfigurable meta materials
\cite{Huang2024-3, GutierrezPrieto2024}, or even gaining insights into
quick reaction mechanisms observed in nature \cite{Forterre2005,
  Smith2011, Forterre2013}.

\begin{figure}[t]
  \centering
  \subfloat[\label{fig:1a}]{\includegraphics[width=0.49\textwidth]{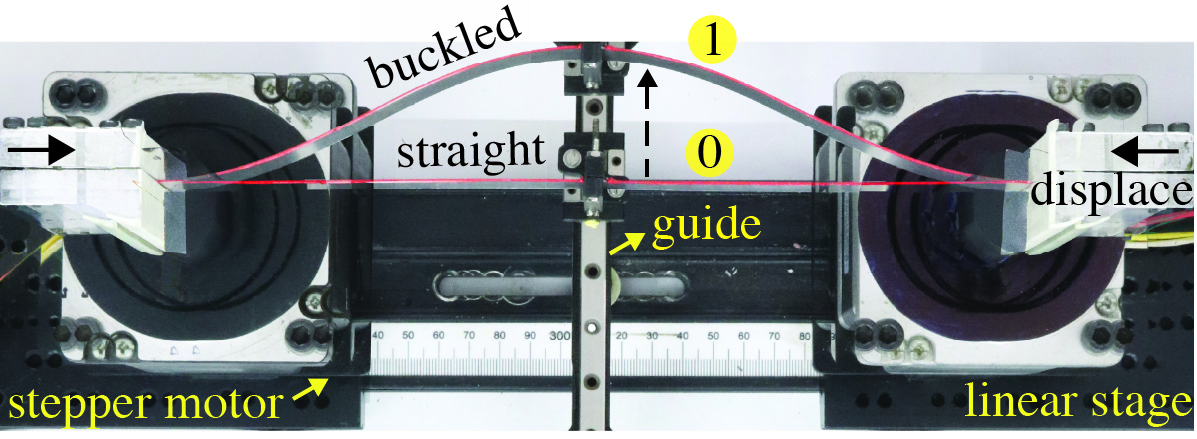}}
  \hfill
  \subfloat[\label{fig:1b}]{\includegraphics[width=0.49\textwidth]{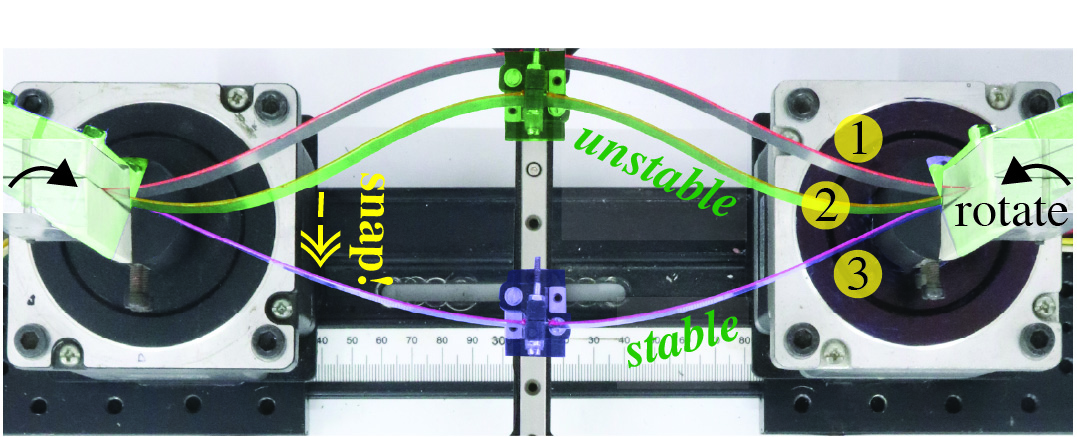}}
  \caption{The dynamic snap-through problem studied in this work. A
    planar elastic arch with a straight unstressed configuration is
    buckled as shown in (a). Then, rotating the ends quasistatically
    as shown in (b) drives the arch to a limit point, when the
    structure snaps dynamically to a distant equilibrium.  The central
    guide rail constrains the arch to deform symmetrically
    throughout.}
  \label{fig:1}
\end{figure}

The specific problem we study here is motivated by a novel application
in which an elastic ribbon with actuated ends snaps underwater. The
fluid's reaction to the instability causes the ribbon to propel in the
direction opposite its snapping motion as discussed in \cref{sec:2}.
Analyzing the propulsion and flow characteristics demands quantifying
the ribbon's rapid shape transition during the snap. In particular,
the critical actuation to instigate the instability, and the ribbon's
pre- and post-snap profiles alone, do not suffice to study the
fluid-structure interaction. With this, and related applications
requiring details of the transient behaviors of snapping structures as
context, we consider a simpler scenario of the instability in a planar
arch.

Our work here focuses on the problem depicted in \cref{fig:1}. The
figure shows a straight elastica compressed past its buckling load and
held with rotatable clamps. When the clamps are turned to a critical
angle, the arch spontaneously snaps from an unstable
equilibrium. Constraining the midpoint of the structure to a guide
ensures that the deformation remains symmetric throughout.  Notice
that, unlike an arch snapping under a transverse load, this problem is
displacement (more correctly, angle) controlled. Before reaching the
snap angle, the dependence of the elastic energy on the arch
midpoint's location has a pair of minima separated by a maximum,
revealing the existence of three equilibria. At the critical clamp
angle, a pair of unstable and stable equilibria annihilate, resulting
in a saddle-node bifurcation that triggers the snap-through
instability. Our main contribution in this article is a single
degree-of-freedom (dof) model describing this dynamic phenomenon and a
detailed examination of the model's predictions.

Variants of the snapping problem in \cref{fig:1} have been examined in
the literature. A majority of them are restricted to the quasi-static
setting and share the goal of determining stability loss at the
critical point and computing vibration frequencies at the snapped
configuration \cite{Chen2008, Zhao2008, Plaut2009, Beharic2014,
  Wan2020}. More recent studies have explored the transient nature of
the phenomenon. The work in \cite{Gomez2016} provides insights on
snap-through solutions near the unstable configuration to explain the
transient behavior observed at the onset of the snap in an
asymmetrically deforming arch.  Normal forms of the governing
equations help discriminate the instability type and study the
incipient dynamics; \cite{Radisson2023-2} provides a procedure for
deriving them. Unlike these studies, our work here seeks to model the
dynamic transition over the entire snap duration, rather than just
near the critical point.

It is important to note the overwhelming dominance of linearized
kinematics in studying problems of arch (in)stability. Linearization
simplifies the elastica equations to the technical beam theory. Both
studies \cite{Gomez2016, Radisson2023-2} noted above, for instance,
leverage the linearity of the Euler-Bernoulli beam equation in their
analyses of dynamic solutions. Characterizing the locus of critical
points for snap-through with the elastica theory is, expectedly, more
challenging \cite{Cazzolli2019}.
Crucially, linearization affords the benefit of adopting modal
expansions for solutions \cite{Harvey2015, Hussein2020}. The utility
of these expansions is particularly evident in addressing design
problems, such as in devising compliant mechanisms with guaranteed
mono- or bi-stability \cite{Palathingal2018}, or say, pre-shaping
arches to yield a desired force-displacement response
\cite{Palathingal2017}. In the dynamic setting, resorting to modal
expansions with time-dependent coefficients provides a systematic
procedure to transform the PDEs governing the arch's response to a
finite set of ODEs for the (generalized) degrees of freedom. However,
modal solutions are not meaningful in studying snap-through at a
saddle-node bifurcation, which is the case in our problem. To wit,
expansions using modes computed at the stable configuration are useful
in predicting post-snap vibrations. During the course of the snap,
however, modes at neither the unstable nor the stable state are
physically appropriate. In this regard, reduced order models can yield
computationally efficient solutions with good accuracy by diligently
constructing problem-specific basis sets for solution approximation
\cite{Howcroft2018, deBono2024}.

Our work here does not propose a toy model for the problem in
\cref{fig:1}. Such models are useful in their own right to gain
qualitative insights.  For instance, a simple von Mises truss-type
model consisting of a pair of oblique linear springs, a torsion spring
and a lumped mass suffices to reproduce the symmetric snap-through
instability studied here, cf. \cite{wiebe2013characterizing,
  Wang2024}.


Our approach to modeling the snap-through dynamics of the elastica
does not rely on asymptotic expansions or normal forms of the
governing equations, linearizing the arch's kinematics or constructing
modal solutions. Instead, the model is based on a serendipitous
observation from examining arch profiles recorded with a high-speed
camera during experimental trials using the setup in \cref{fig:1}. We
found that the instantaneous profiles in the frames closely resembled
those observed in a quasi-static displacement-controlled test
performed to measure the arch's force response to an imposed midpoint
deflection.  Although the boundary conditions at the clamps and the
symmetry constraint at the center coincide in the two experiments, the
observation is surprising.  There is, after all, little justification
to expect arch shapes during a force-free dynamic snap to be
correlated with those manifested in a forced geometrically-constrained
quasistatic test. Further experimental trials and extensive finite
element (FE) simulations corroborated this coincidence over a wide
range of arch depths. This phenomenological observation, alongside
energy conservation, constitutes the essence of our model. We discuss
its detailed formulation in \cref{sec:3}.

The utility of the proposed model depends foremost on its prediction
accuracy. We examine this aspect in detail in \cref{sec:4}. Besides
validating its solutions by comparisons with experiments and FE
simulations, the model also reveals interesting features of the
snapping phenomenon; we highlight these in \cref{sec:5}. A second
question concerns the purpose served by the model in light of the
well-established dynamical theory of the elastica \cite{Caflisch1984,
  Patrcio1998}. Indeed, a broad class of geometrically nonlinear beam
theories is well-suited to model the problem we study \cite{Simo1985,
  Meier2017}. Distinctions between them, usually stemming from details
in the treatment of (in)extensibility and (un)shearability, are
insignificant in the present context because the arch is slender and
its deformation remains bending-dominated. Computer implementations of
these beam models are widely available in general-purpose FE codes.
In fact, we rely on ABAQUS simulations to evaluate the proposed
model's accuracy. However, we draw attention to challenges that
persist, especially for simulating dynamic snap-through
instabilities. Numerical instabilities and non-convergence often
plague simulations despite employing time integrators with guaranteed
\emph{linear} stability.  The large accelerations encountered in the
problem necessitate adaptive time stepping, often leading to
undesirably small step sizes \cite{Chandra2013,
  Chandra2015}. Numerical dissipation caused by non-physical
high-frequency oscillations can result in unacceptably large energetic
deviations \cite{Kuhl1996}.  Our experience with the FE simulations
required in the validation studies in \cref{sec:4} also affirms
reports in the literature that FE simulations of dynamic snap-through
are far from automatic and seldom robust \cite{Chandra2012}. The
proposed model is, then, a compelling alternative. As a single dof
model, the computational benefits it offers over conventional
numerical simulations require little elaboration. The model only
requires the solution of ODEs to determine arch profiles during the
snap-through, and can even leverage closed-form solutions in the
elastica theory to render it extremely efficient. Furthermore, the
model preserves energy exactly and hence enjoys guaranteed
stability. The proposed model can serve, for instance, iterative
design studies of a soft robot exploiting snap-through instabilities
\cite{Tong2025}, for which relying on dynamic FE simulations is likely
impractical.

The remainder of the article is organized as follows. We begin in
\cref{sec:2} with a concise discussion of an application exploiting a
snap-through instability for swimming.  The application motivates the
problem of dynamic snap-through of a buckled elastic arch that we
study.  We formulate a model for the problem in \cref{sec:3} and
examine its predictive accuracy in detail in \cref{sec:4}.  We discuss
features of the model and record observations on the snap-through
phenomenon based on its predictions in \cref{sec:5}. We close with a
summary in \cref{sec:6}.


\section{Swimming by snapping: a motivating application}
\label{sec:2}
The problem in \cref{fig:1} is motivated by an application that
harnesses the rapid energy released by a snapping elastic ribbon for
propulsion underwater \cite{Yamada2011, Chen2018, Chi2022}.  In the
following, we briefly explain the novel working principle underlying
the application, which provides useful context for the subsequent
sections.

\begin{figure}[t]
  \centering
  \subfloat[\label{fig:2a}]{\includegraphics[height=0.3\textwidth]{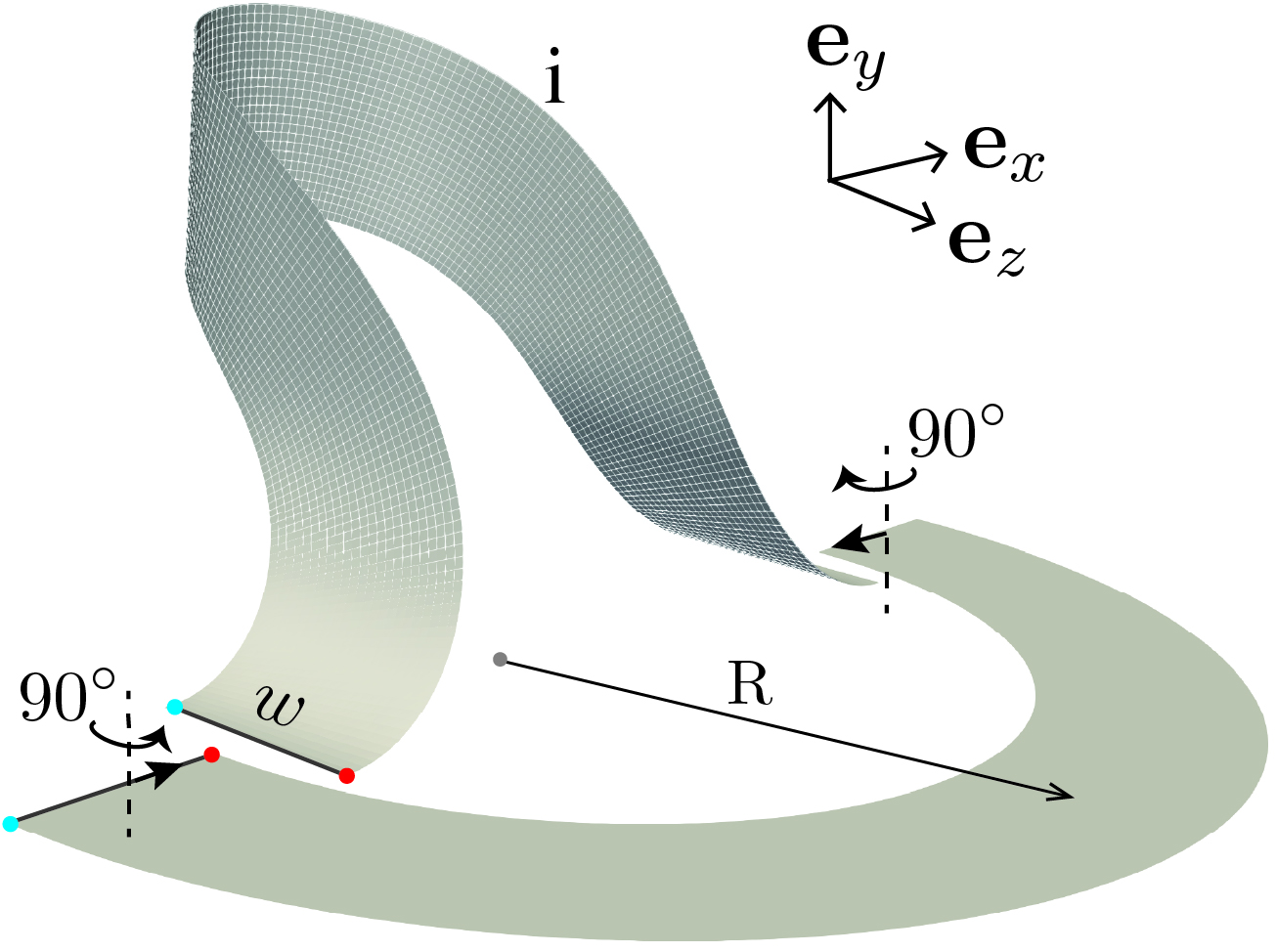}}
  \hfill
  \subfloat[\label{fig:2b}]{\includegraphics[height=0.35\textwidth]{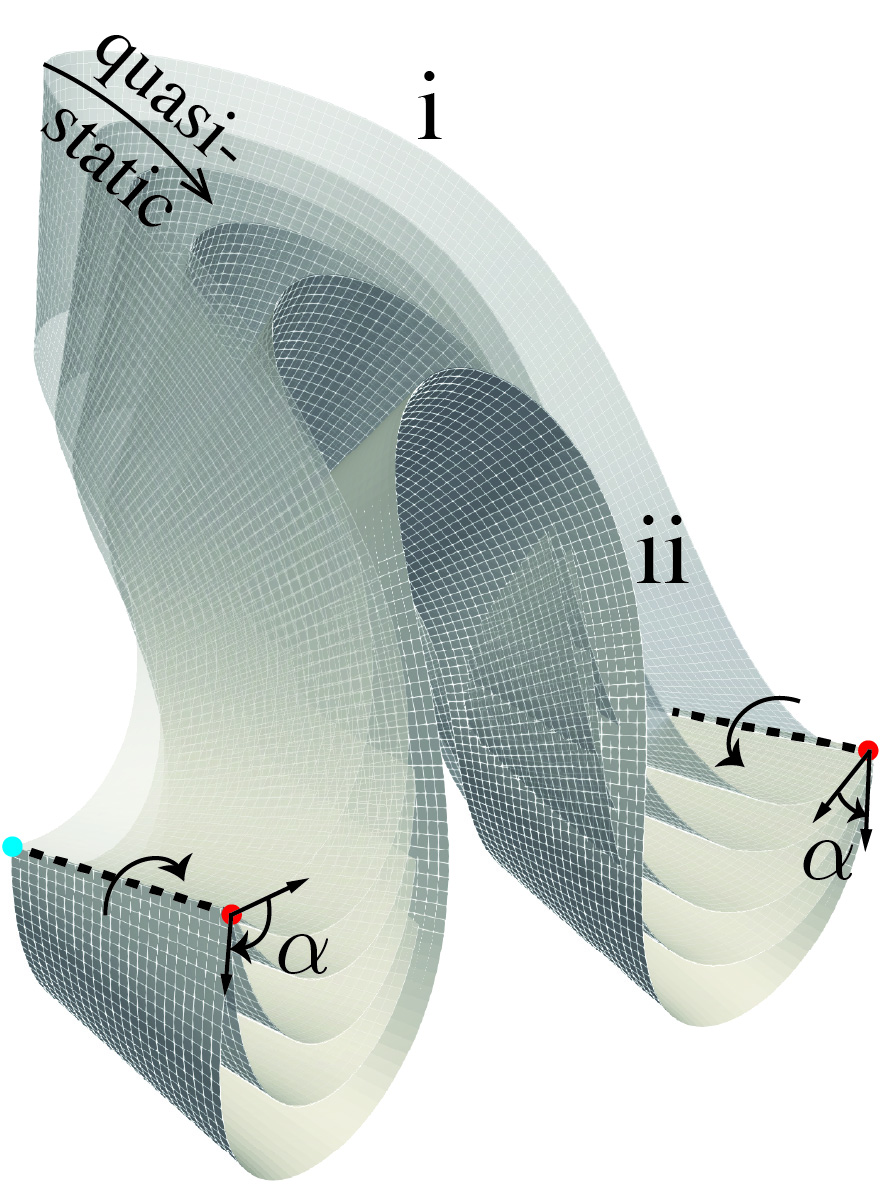}}
  \hfill
  \subfloat[\label{fig:2c}]{\includegraphics[height=0.35\textwidth]{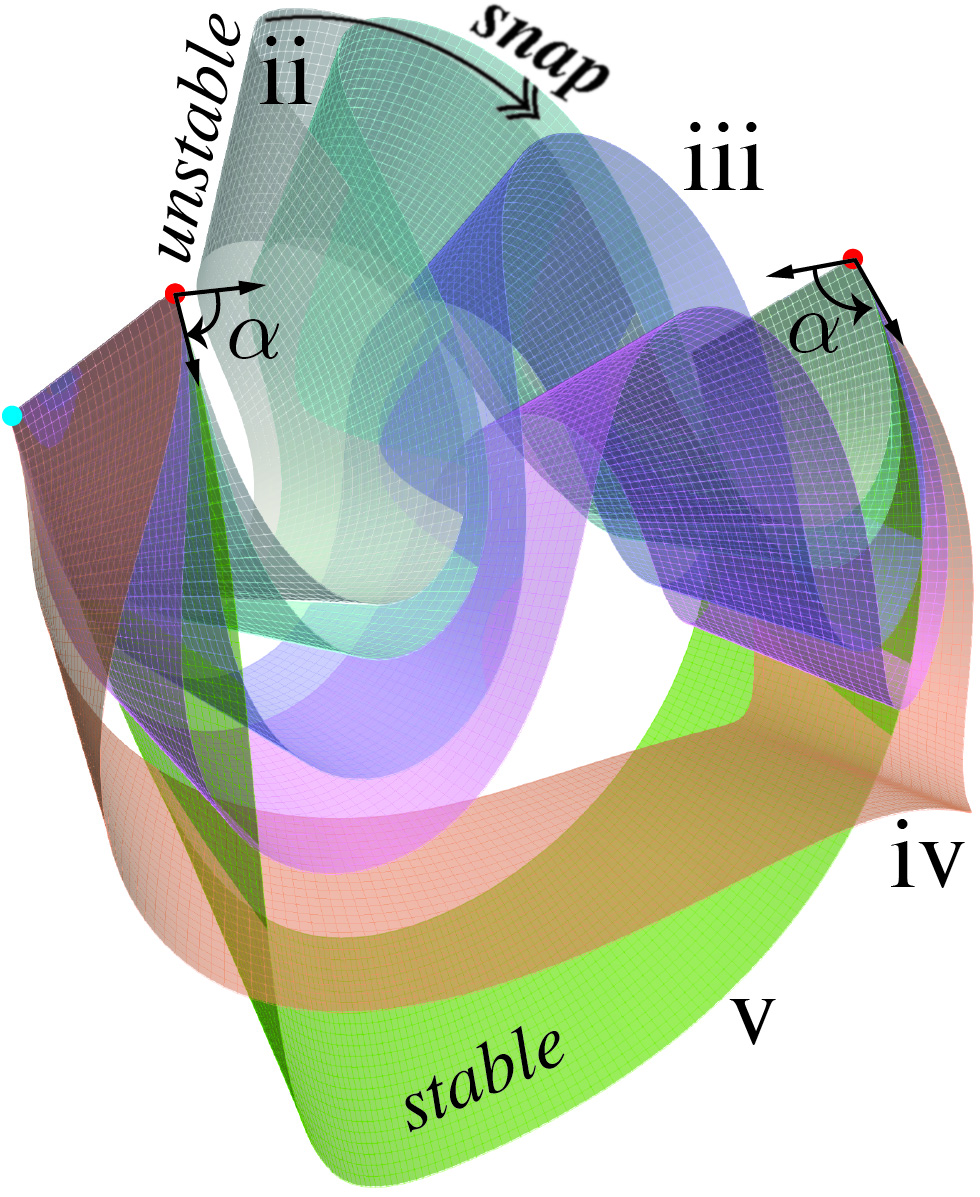}}
  \caption{The application to swimming discussed in \cref{sec:2}
    exploits the snap-through of a semi-annular elastic ribbon. The
    ribbon is pre-buckled to state i shown in (a). Its ends are
    gradually rotated until it reaches the unstable configuration $ii$
    in (b). Thereafter, the ribbon spontaneously snaps to $v$ as
    depicted in (c). Ribbon configurations in (a,b) were computed
    using quasistatic FE simulations, while those in (c) employed a
    dynamic simulation.}
 \label{fig:2}
\end{figure}

\begin{figure}[t]
  \centering
  \subfloat[\label{fig:3a}]{\includegraphics[width=0.48\textwidth]{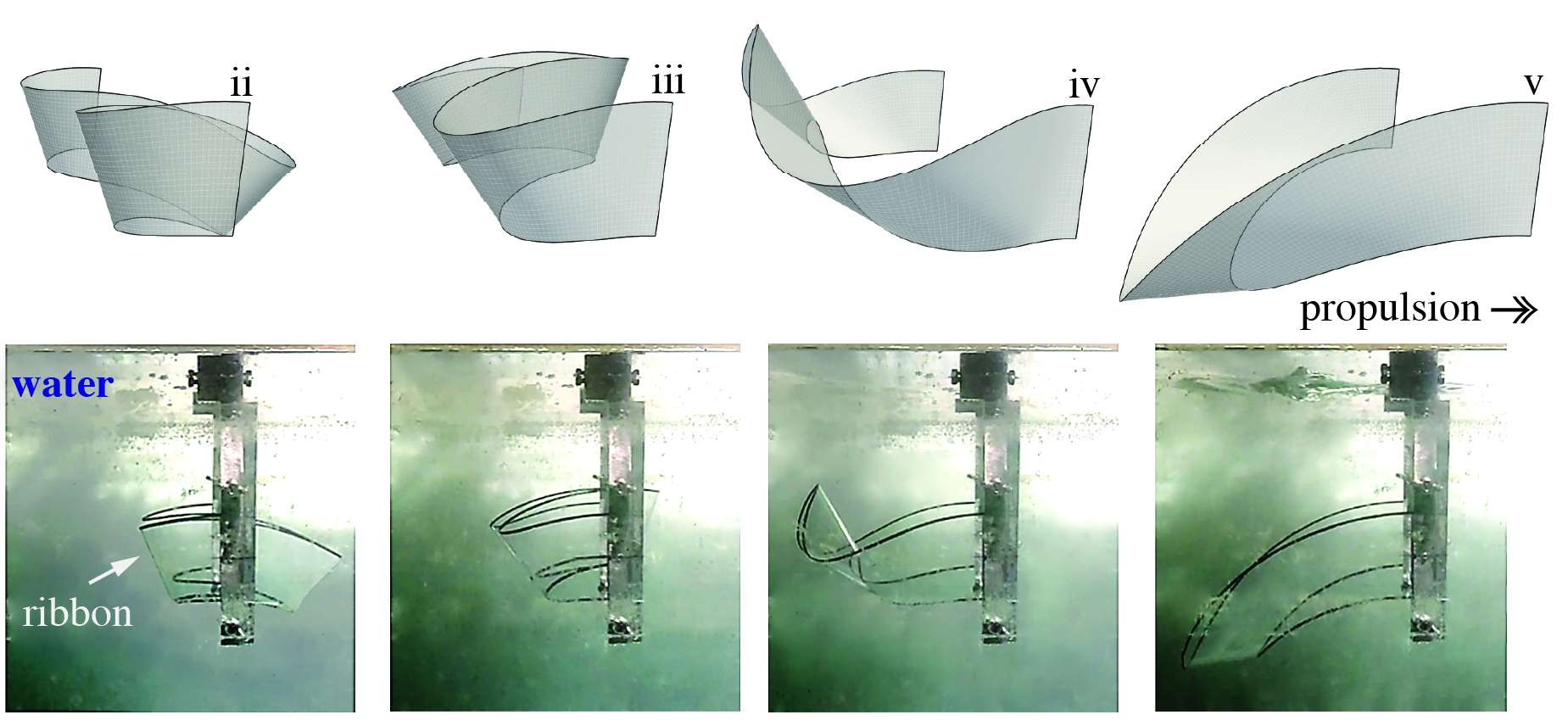}}
  \hfill
  \subfloat[\label{fig:3b}]{\includegraphics[width=0.52\textwidth]{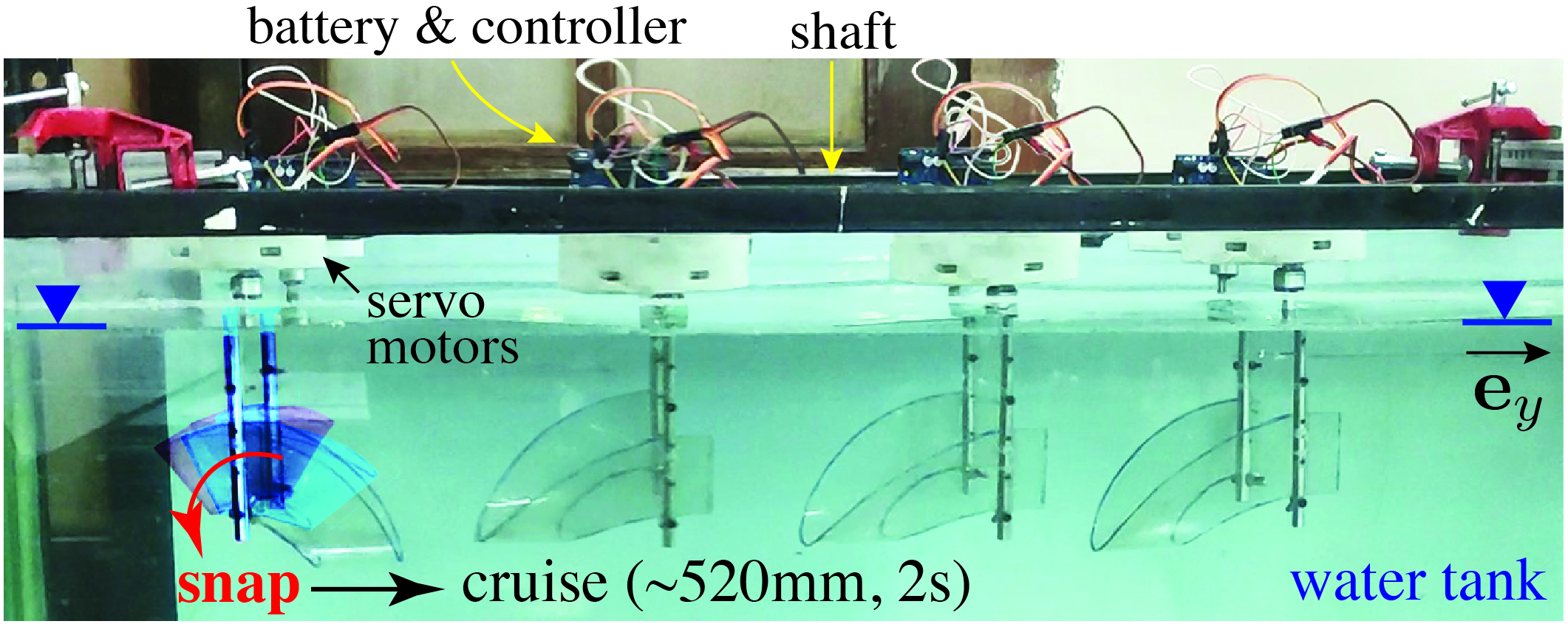}}
  \\
  \subfloat[\label{fig:3c}]{\includegraphics[height=0.26\textwidth]{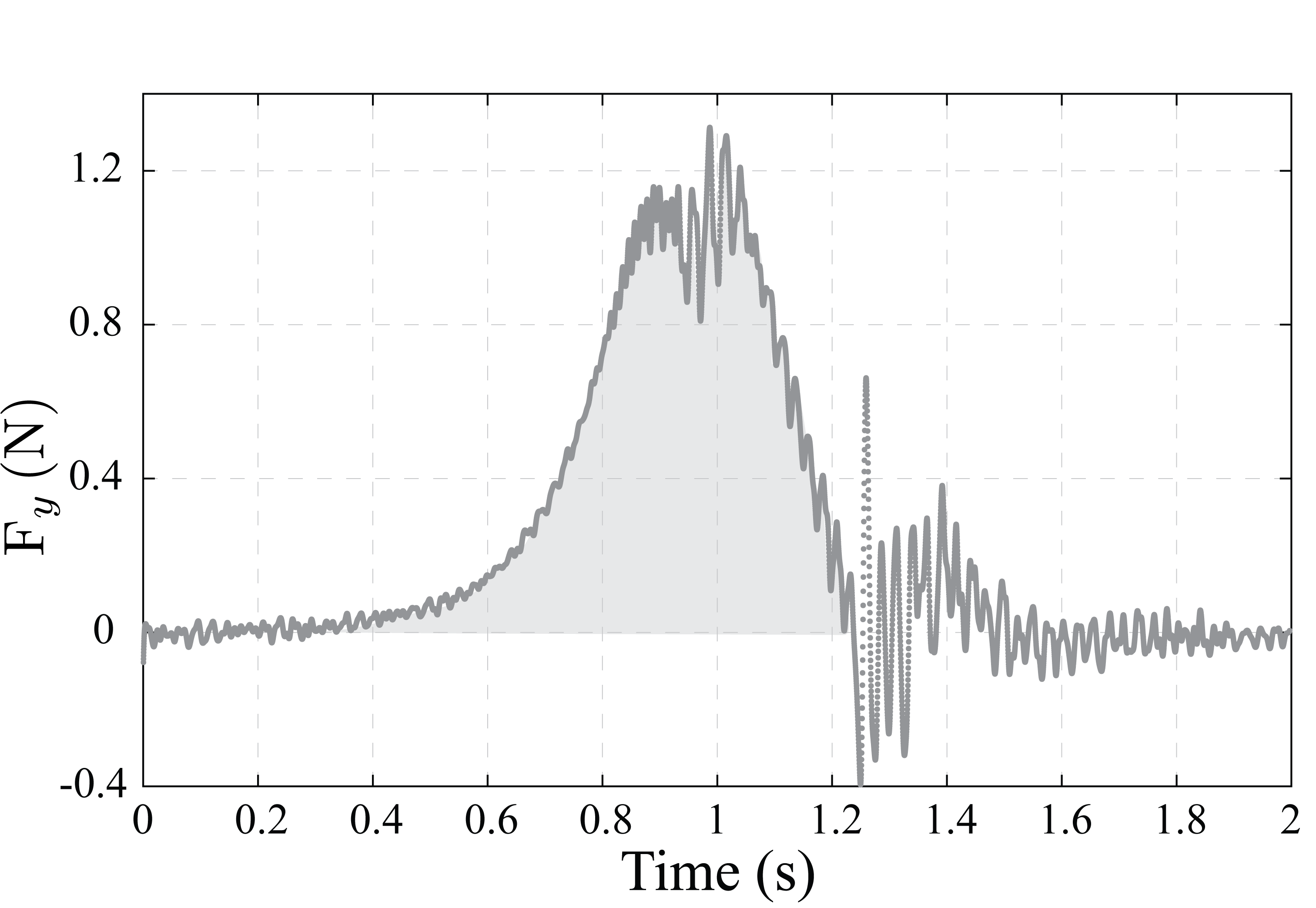}}
  \hfill
  \subfloat[\label{fig:3d}]{\includegraphics[height=0.26\textwidth]{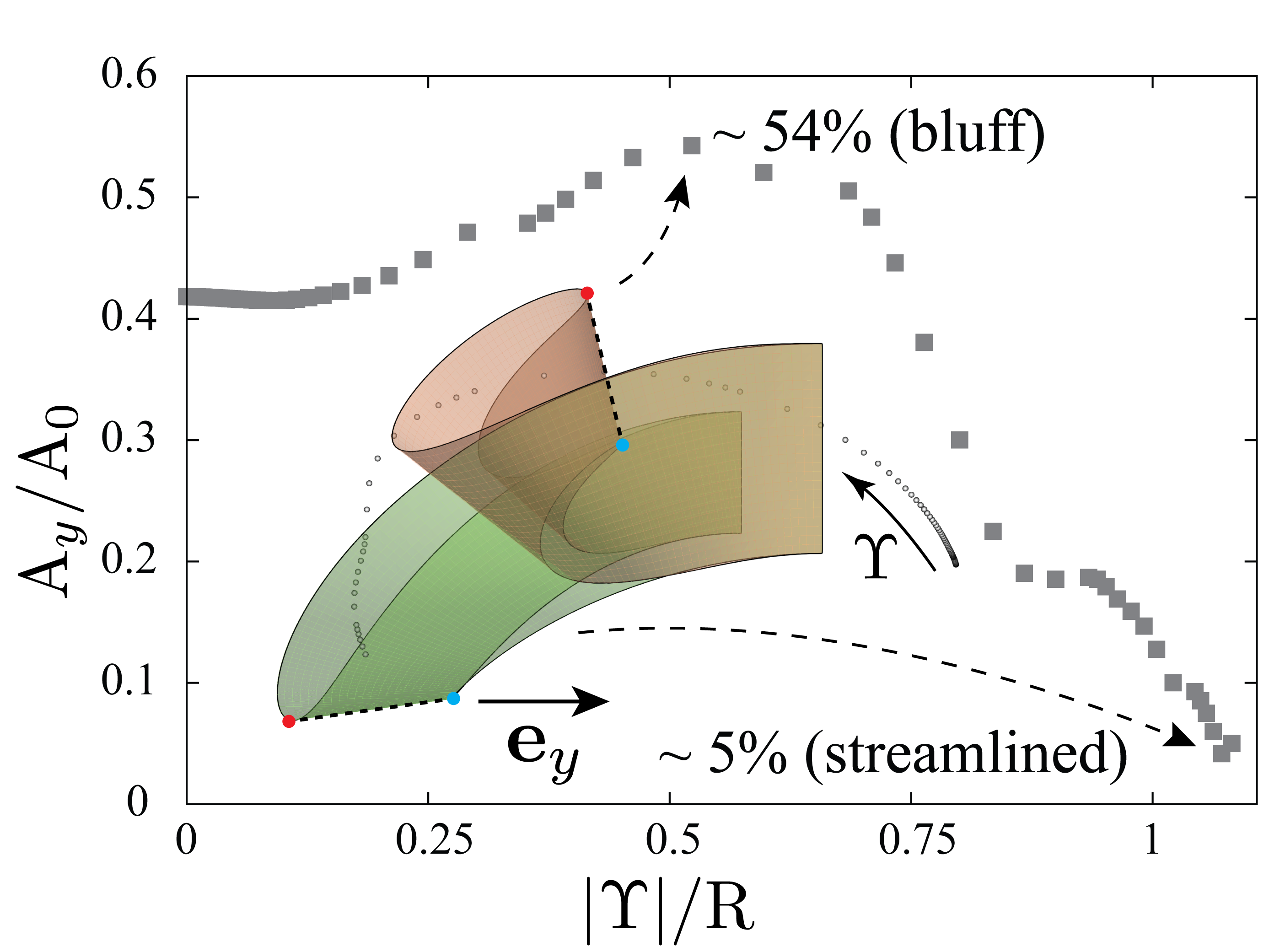}}
  \hfill
  \subfloat[\label{fig:3e}]{\includegraphics[height=0.26\textwidth]{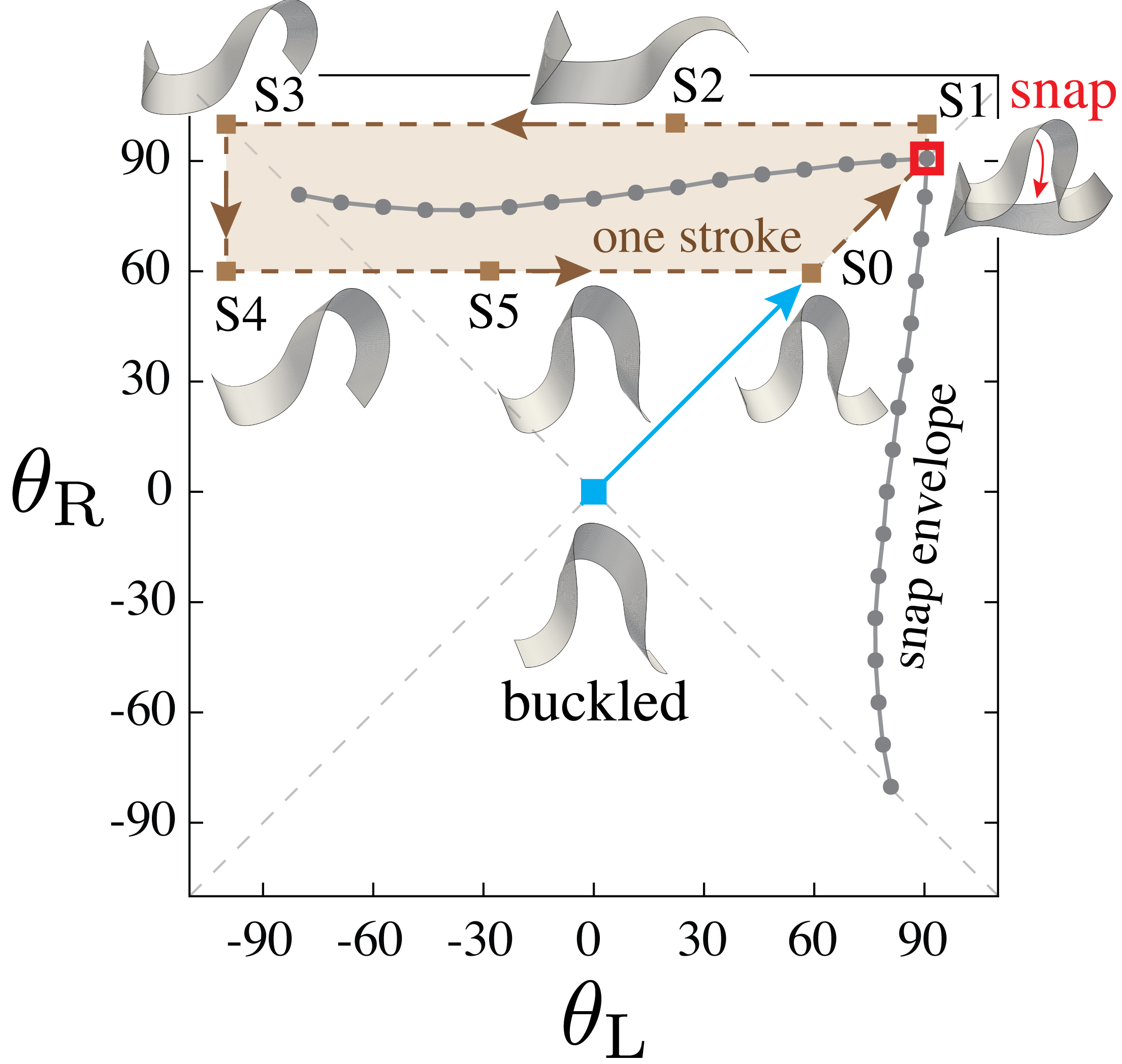}}
  \caption{(a) Snapshots of a ribbon snapping underwater, highlighting
    similarities with the simulated profiles shown in \cref{fig:2}.
    (b) A tether-less prototype that propels underwater by harnessing
    the snap-through in (a). (c) Experimental measurement of the
    propulsive component of the reaction force using the setup in (a)
    with a polycarbonate ribbon. (d) Area the ribbon's surface projected on
    the plane orthogonal to the swim direction during its snap from
    the simulation in \cref{fig:2} shows that the shape transition
    during the snap ensures a large displaced volume at the start of
    the snap and a streamlined profile at the end. (e) The snap
    envelope helps plan the end actuation such that ribbon snaps just
    once per stroke.}
    \label{fig:3}
\end{figure}

\noindent
\paragraph{Working principle.}
\Cref{fig:2} shows a semi-annular elastic ribbon of radius ${\rm R}$
and width $w$, with $w/{\rm R}\approx 0.38$. The ribbon is centered at
the origin and contained in the ${\bf e}_x-{\bf e}_y$ plane.
Retracting its diametrical edges towards the center along the
${\bf e}_x$ direction causes the ribbon to buckle out of plane. Then,
we rotate each edge by $90^\circ$ about an axis parallel to
${\bf e}_y$ as indicated in \cref{fig:2a}. The ribbon assumes a
symmetric $\Omega$-shaped three-dimensional equilibrium configuration
labeled `i' with its short edges aligned along the ${\bf e}_z$
direction.  Now, gradually rotating each edge about the ${\bf e}_z$
axis causes the structure to deform quasistatically as shown in
\cref{fig:2b}. Eventually, the ribbon reaches the unstable
configuration labeled `ii' in the figure. At this limit point, the
ribbon snaps dynamically to the state `v' indicated in \cref{fig:2c}.
The intermediate shapes `iii' and `iv' indicated in the figure are
manifested during the snap, but are not equilibrium configurations.
During the snap, the ribbon's midpoint traverses a curvilinear
trajectory $\Upsilon$ in the ${\bf e}_y-{\bf e}_z$ plane (see
\cref{fig:3d}). The application to swimming exploits the spontaneous
snap-through in \cref{fig:2c}. When immersed underwater, the transient
motion of the ribbon during the snap displaces the ambient fluid. The
reaction exerted by the fluid on the ribbon manifests as a net force
on the snapped structure; the component ${\rm F}_y$ of this force
propels the ribbon along the ${\bf e}_y$ direction.

\noindent
\paragraph{Simulations and underwater tests.}
The kinematics of the ribbon depicted in \cref{fig:2} are the result
of FE simulations employing geometrically nonlinear plate elements
(S4R) in ABAQUS. The simulation uses a combination of static and
dynamic phases, with the latter being restricted to the snapping
duration. However, the simulations ignore the fluid-structure
interaction. Hence, the computations provide an account of the
ribbon's shape transformation sequence but not the propulsive forces.

To assess the propulsion, we resort to an experimental measurement.
\Cref{fig:3a} shows a physical prototype of the ribbon having
dimensions ${\rm R}=124\,{\rm mm}, w=47\,{\rm mm}$ cut from a
polycarbonate sheet of thickness $0.5\,{\rm mm}$. The straight edges
of the ribbon are attached to the shafts of a pair of servo
motors. The fixture supporting the motors is in turn rigidly coupled
to a stationary load cell. The ribbon is submerged underwater in a
small laboratory tank, driven to the limit point and driven to snap
just as in \cref{fig:2c}. \Cref{fig:3a} shows the ribbon profiles
recorded during its underwater snap-through. Therein, we highlight the
similarities of these profiles with those shown in \cref{fig:2c} in
the absence of an ambient medium. This observation suggests that the
ribbon is sufficiently stiff to drive the flow, i.e., the ambient
fluid does not significantly influence the ribbon's shape during the
snap.

Next, with the ribbon held submerged underwater, we record the force
history during the snap-through. The measured forces include
contributions from the structure and from the reaction forces exerted
by the fluid on the ribbon during the snapping motion.  \Cref{fig:3c}
shows a representative measurement of the component ${\rm F}_y$ of the
force along ${\bf e}_y$. Notice that ${\rm F}_y$ is positive over the
entire duration of the snap, as is desirable for propulsion.
Post-snap disturbances seen in the force history are difficult to
interpret, especially because they include contributions from fluid
reflections off the walls of the narrow testing tank.

To test the snap-induced propulsion of the ribbon, we detach the
fixture coupling the ribbon-motor assembly to the load cell. The
assembly is augmented with a micro-controller to drive the motors and
a battery for power supply. The resulting tether-less prototype is
mounted on a shaft parallel to the length of the tank $({\bf
  e}_y)$. \Cref{fig:3b} shows snapshots of the ribbon during the snap
and the subsequent translation of the prototype on the shaft. A video
recording of the test is included in the set of supplementary
materials accompanying the article. The ribbon translates a distance
of about $520$ {\rm mm} over approximately $2{\rm s}$, before coming
to rest.

\paragraph{Rationale for ribbon shapes.} The ribbon morphologies
employed to realize the propulsive motion are not coincidental. The
shapes realized during the snap are such that the projected area of
the ribbon's surface on the plane orthogonal to the propulsion
direction is large at the start of the snap, but reduces to a small
fraction at the end. \Cref{fig:3d} illustrates this point by plotting
the fraction of the ribbon's surface area projected on the
${\bf e}_x-{\bf e}_z$ plane for the profiles realized during the
dynamic snap-through simulated in \cref{fig:3a}. We see that the
projected area peaks at about $54\%$ during the snap, and reduces to a
mere $5\%$ at the end of the snap. Consequently, the ribbon elicits a
large reaction force from the fluid due to the mass it displaces
during the initial phase of the snap. At the snapped configuration,
the ribbon presents a streamlined profile, enabling it to cruise along
the length of the tank. Since the snapping simulation does not account
for the fluid-structure interaction, we plot the projected area in
\cref{fig:3d} as a function of the distance $|\Upsilon|$ traversed by
the midpoint of the ribbon, rather than time.  The rotation of the
ribbon's cross-section relative to the propulsion direction can also
be observed in \cref{fig:3d}, where the central radial line is nearly
orthogonal to ${\bf e}_y$ during the snap but turns parallel to it at
the end.  In this context, we mention the work of \cite{Yamada2011}
advocating a twist-induced snapping morphology \cite{Sano2019} for
underwater propulsion. Such an arrangement is well-suited for
achieving quick-turn maneuvers rather than rectilinear motion targeted
in our prototype.

\noindent
\paragraph{Multi-stroke swimming.}
The last aspect of the application we highlight concerns the strategy
devised to achieve multi-stroke swimming. For the ribbon to execute
repeated strokes, it is necessary to revert it from the snapped
configuration to the pre-snap state. It is crucial, however, to
realize this transformation without incurring a snap-back instability
which would reverse the propulsion achieved, resulting in little net
locomotion.
Our configuration-reversal strategy for the ribbon is based on
computing its {snap envelope} as a function of its end rotations
$\theta_{\rm L}$ and $\theta_{\rm R}$. To this end, we use FE
simulations similar to those employed in \cref{fig:2} and identify the
locus of end rotations $\theta_{\rm L}$ and $\theta_{\rm R}$ at which
the ribbon snaps. \Cref{fig:3e} shows the snap-envelope computed this
way.



Each point on the snap locus represents two states of the ribbon--- an
unstable pre-snap and a stable post-snap configuration. Hence, the
snap-through employed in our application is located at the
intersection of this envelope with the $\theta_{\rm L}=\theta_{\rm R}$
line. Observe from \cref{fig:3e} that the snap envelope is an open
curve (the reflection of the curve about the origin defines the snap
envelope for a buckled ribbon having an inverted configuration).
Then, rather than incur a snap-back instability by reversing course
along the $\theta_{\rm L}=\theta_{\rm R}$ line, we plan a path in the
$\theta_{\rm L}-\theta_{\rm R}$ plane that circumvents the envelope
altogether. The closed contour
${\rm S1\rightarrow S2\rightarrow S3\rightarrow S4\rightarrow
  S5\rightarrow S0\rightarrow S1}$ indicated in the figure defines a
single stroke for the ribbon. The path has the crucial property that
it crosses the snap envelope just once, in the
${\rm S0\rightarrow S1}$ segment. Ribbon profiles are symmetric over
the ${\rm S0\rightarrow S1}$ segment, and asymmetric elsewhere. The
actuation sequence over the entire path is quasistatic; the ribbon's
response is quasistatic as well, except for the snap-through. In
practice, we find that the ribbon ends can be actuated reasonably
quickly without inducing appreciable ambient flow, so that only the
snap-through contributes to propelling the ribbon.


\noindent
\paragraph{Discussion.}The application of a snapping ribbon to
swimming naturally opens up numerous avenues for investigation, from
locomotion efficiency to achieving curvilinear motions by choosing
symmetry-breaking snap-through configurations suggested by
\cref{fig:3e}. However, our sole purpose is to highlight the parallels
between the application and the snap-through problem studied next. The
arch in \cref{fig:1} and the ribbon in \cref{fig:2} are both buckled
to symmetric configurations, driven to an instability at a turning
point through incremental end rotations, and dynamically jump to a
distant equilibrium configuration. In essence, the snapping elastica
problem we study is a planar version of the three-dimensional
phenomenon underlying the swimming application. The only other
distinction concerns the symmetry of the deformation- while a
sufficiently wide ribbon naturally maintains symmetry when snapping,
this is not guaranteed in the case of a planar arch. Instead, we
impose symmetry in our problem as a constraint. The purpose of the
model proposed to describe the dynamics of a snapping arch is also
amply justified by the swimming application. In particular, estimating
the energy difference between the pre- and post-snap configurations
alone does not suffice; the propulsive force depends crucially on the
evolution of the ribbon's shape during the phenomenon.





\section{Model for dynamic snap-through}
\label{sec:3}
\begin{figure}[t]
  \centering
  \subfloat[\label{fig:4a}]{\includegraphics[height=0.21\textwidth]{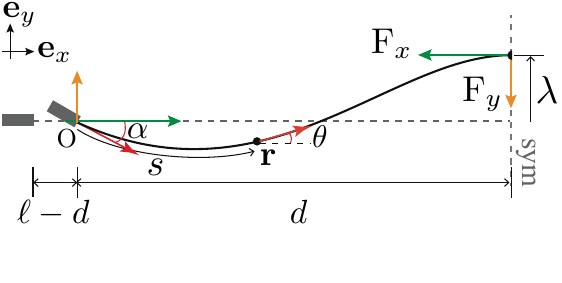}}
  \hfill
  \subfloat[\label{fig:4b}]{\includegraphics[height=0.21\textwidth]{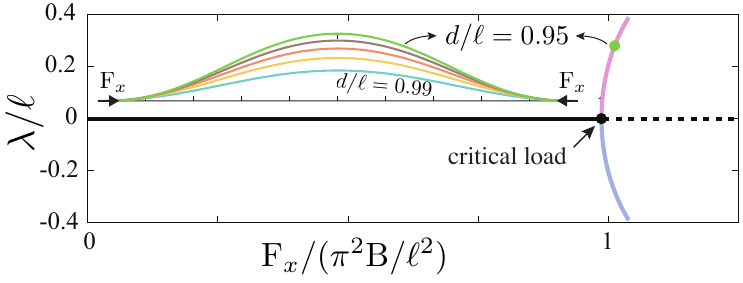}}
  \caption{(a) Coordinate system and notation used in the discussion
    of equilibrium solutions of the elastica in \cref{sec:3-1}. Due to
    the reflection symmetry assumed, only half the span is shown. (b)
    Bifurcation diagram and arch profiles for the buckling problem.}
  \label{fig:4}
\end{figure}

We devote this section to formulating a model for the dynamic
snap-through problem of the planar elastica in \cref{fig:1}.  We
assume that the elastica has length $2\ell$ and is straight in the
unstressed state.  Noting the reflection-symmetry imposed by the
guideway midspan, we restrict attention to a symmetric half of the
structure. Accordingly, we introduce the notation and coordinate
system required in the remainder of the discussion in
\cref{fig:4a}. The arc length parameter along the centerline is
denoted by $s$ and time by $t$. We frequently use the shorthands
$(\cdot)'\equiv \partial(\cdot)/\partial s$ and
$\dot{(\cdot)} \equiv \partial(\cdot)/\partial t$ to denote
derivatives with respect to the two parameters.

Since the arch is slender and we expect strains to remain small, we
assume the centerline of the elastica to be inextensible and invoke a
linear constitutive relationship. As a consequence of the former, the
kinematics of the arch is conveniently described using its tangent
inclinations. Given the inclination angles $s\mapsto \theta(s)$, the
centerline follows as
${\bf r}(s) = \int_{0}^s(\cos\theta(\sigma),\sin(\sigma))\,d\sigma$
which is a curve parameterized by arc length. In particular, the half
span $d$ and the height $\lambda$ of the arch are given by
\begin{align*}
  (d,\lambda) \equiv {\bf r}(\ell) =
  \int_{0}^\ell(\cos\theta(\sigma),\sin(\sigma))\,d\sigma.
\end{align*}
The proportionality factor in the moment-curvature relationship for
the arch is the bending modulus ${\rm B}$.
Throughout our presentation, we assume it to be uniform, i.e.,
independent of $s$. This is the case in our experiments performed with
arches having uniform rectangular cross sections, and in all the
numerical simulations shown.

\subsection{Equilibrium states}
\label{sec:3-1}
It is instructive to examine the pre- and post-snap equilibrium
configurations of the elastica before proceeding to the dynamic
transition between them. Of direct concern to us are three problems---
buckled solutions without end rotation, solutions with end rotations
realized prior to and after snapping, and an auxiliary problem we
introduce for subsequent use.

\noindent
\paragraph{Buckled solutions.} These describe the deflected
equilibrium configurations of the arch depicted in \cref{fig:1a}, as
the ends are compressed to realize a desired half-span $d$. The clamp
orientations are horizontal. Since the midpoint of the arch slides
freely on the guide, the reactions at the clamps are purely
horizontal.  Thus in \cref{fig:4a}, we set ${\rm F}_y=0$ and
$\alpha=0$.  The statement of moment balance in terms of the tangent
inclination $s\mapsto \eta(s)$ is given by
${\rm B}\eta'(s) = \int_{s}^\ell{\rm F}_x\sin\eta(\sigma)\,d\sigma$.
Differentiating it with respect to $s$, we arrive at the boundary
value problem for $\eta$:
\begin{align}
  \begin{cases}
    &\eta''(s) + {\rm F}_x\sin\eta(s)=0 \\
    &\eta(s=0)=\eta(s=\ell)=0,
  \end{cases}
      \label{eq:3-2}
\end{align}
where the boundary conditions are consequences of the horizontal
orientation of the clamp at the left end and the symmetry imposed
midspan.

As seen in \cref{fig:4b}, problem \cref{eq:3-2} has a trivial solution
$\eta=0$ until ${\rm F}_x$ exceeds the critical load
$\pi^2{\rm B}/\ell^2$ corresponding to a supercritical pitchfork
bifurcation. Thereafter, the trivial branch is unstable and the
elastica buckles to a bent profile. For definiteness, we follow the
branch with $\lambda>0$ and proceed until $d$ equals a desired
fraction of $\ell$. The figure shows buckled profiles realized for
$d/\ell = 0.99,0.98,\ldots,0.95$. The buckled arch in \cref{fig:1a}
corresponds to the case $d/\ell=0.95$.

\noindent
\paragraph{Pre- and post-snap solutions.} Next, we quasistatically
rotate the clamps while permitting the midpoint of the arch to slide
freely along the vertical guide. The reaction force at the clamp is
again parallel to ${\bf e}_x$. Its magnitude, however, is implicitly
determined by the constraint that the half-span remain $d$.
Equilibrium solutions $s\mapsto \varphi_{\alpha}(s)$ with clamp angle
set to $\alpha$ hence satisfy:
\begin{align}
  \begin{cases}
    &\varphi_{\alpha}''(s) + {\rm F}_x\sin\varphi_{\alpha}(s)=0 \\
    &\varphi_{\alpha}(0)=-\alpha, ~\varphi_\alpha(\ell)=0,~\text{and}~\int_{0}^\ell\cos\varphi_\alpha(s)\,ds=d.
  \end{cases}
      \label{eq:3-3}
\end{align}

\begin{figure}[t]
  \centering
  \subfloat[\label{fig:5a}]{\includegraphics[height=0.295\textwidth]{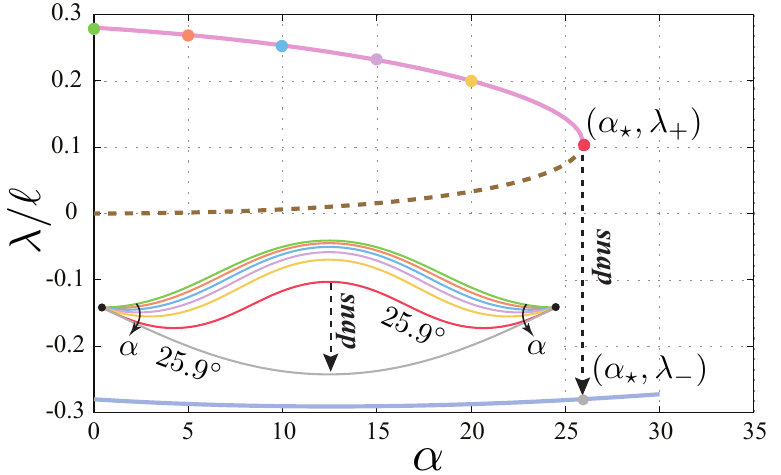}}
  \hfill
  \subfloat[\label{fig:5b}]{\includegraphics[height=0.295\textwidth]{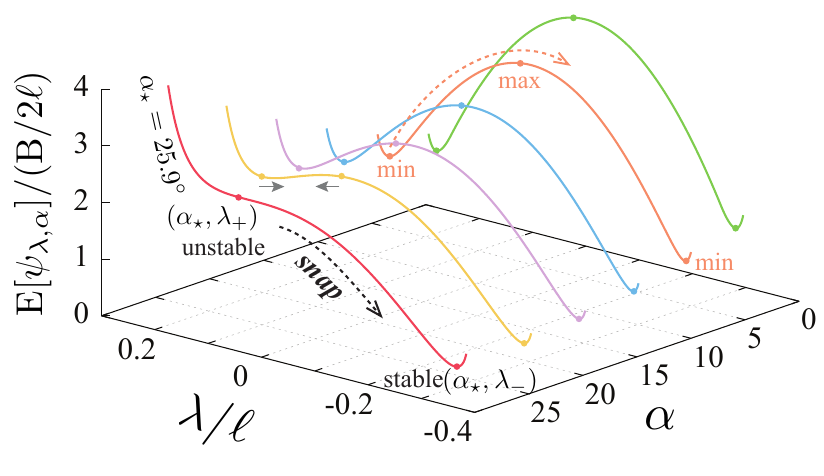}}
  \caption{(a) Bifurcation diagram for problem \ref{eq:3-3}. The
    buckled solution at $\alpha=0$ follows the upper stable branch
    until the critical angle $\alpha_\star$. There, the arch jumps to
    the lower stable branch. Examining the energies of solutions to
    problem \ref{eq:3-4} in (b) confirms the fold bifurcation at
    $\alpha_\star$.}
  \label{fig:5}
\end{figure}

\Cref{fig:5a} shows the bifurcation diagram computed for \cref{eq:3-3}
with $d/\ell=0.95$. The plot reveals that for $\alpha$ ranging from
$0$ to a critical angle $\alpha_\star\approx 25.9^\circ$,
\cref{eq:3-3} has three equilibrium solutions--- a pair of which are
stable and a third which is not. Starting from the buckled solution at
$\alpha=0$, gradually rotating the clamps as done in \cref{fig:1b}
leads the solution quasi-statically along the stable upper branch for
which intermediate arch profiles are shown in the inset of
\cref{fig:5a}. The height of the arch decreases monotonically in the
process. For $\alpha>\alpha_\star$, there is a lone stable equilibrium
branch. Hence, the solution spontaneously switches from the upper to
the lower stable branch at $\alpha_\star$; the arch height $\lambda$
jumps from $\lambda_+$ to $\lambda_-$.
The vertical dashed line in the plot connecting the unstable and
stable solutions conveys that the arch jumps with the clamp angle
remaining constant at $\alpha_\star$. The transition between branches
is dynamic and no point on the line should be interpreted to be an
equilibrium state.

\noindent
\paragraph{An auxiliary problem.} We introduce an auxiliary problem
that aids in formulating the proposed model.  Given parameters
$\alpha, d$ and $\lambda$, we seek equilibrium solutions
$s\mapsto \psi_{\lambda,\alpha}(s)$ satisfying
\begin{align}
  \begin{cases} &\psi_{\lambda,\alpha}''(s) + {\rm F}_x\sin\psi_{\lambda,\alpha}(s) -
    {\rm F}_y\cos\psi_{\lambda,\alpha}(s) = 0, \\
    &\psi_{\lambda,\alpha}(0)=-\alpha,\psi_{\lambda,\alpha}(\ell)=0,
    ~\text{and} \\
    &\int_{0}^\ell(\cos\psi_{\lambda,\alpha}(s),\sin\psi_{\lambda,\alpha}(s))\,ds
    = (d,\lambda),
  \end{cases}
      \label{eq:3-4}
\end{align}
where the reaction forces ${\rm F}_x$ and ${\rm F}_y$ are determined
as part of the solution to satisfy the constraints imposed on the arch
span and height. Solutions of \cref{eq:3-4} help verify that the
critical point at $\alpha=\alpha_\star$ in \cref{eq:3-3} is indeed a
fold bifurcation. To this end, we examine the elastic energies
\begin{align}
  {\rm E}[\psi_{\lambda,\alpha}] \equiv \int_0^\ell {\rm
  B}\psi_{\lambda,\alpha}'^2(s)\,ds. \label{eq:3-5}
\end{align}
For the case $d/\ell=0.95$, \cref{fig:5b} plots profiles of
$\lambda\mapsto {\rm E}[\psi_{\lambda,\alpha}]$ at a few discrete
values of $\alpha$.  Notice that at $0<\alpha<\alpha_\star$, each
energy profile shows three extrema--- a pair of minima and a maximum
sandwiched between them. At these extrema, the vanishing derivative
$\partial{\rm E}/\partial\lambda$ implies ${\rm F}_y=0$, implying that
the corresponding equilibria coincide with solutions of \cref{eq:3-3}.
Thus, the loci of the extrema in \cref{fig:5b} are precisely the
equilibrium branches traced in \cref{fig:5a}. The two stable branches
in the \cref{fig:5a} track the pair of minima in \cref{fig:5b}, while
the unstable branch tracks the maximum.

This relationship between the solution branches in \cref{fig:5a} and
the energy profiles in \cref{fig:5b} confirms the intuitive
expectation that it is possible for the arch to switch between stable
branches for $\alpha<\alpha_\star$ in problem \cref{eq:3-3}. However,
doing so requires overcoming a large energy barrier, and hence a large
external perturbation. Furthermore, notice in \cref{fig:5b} that the
minimum falling in the region $\lambda>0$ and the maximum get
progressively closer with increasing $\alpha$. The stable and unstable
branches converging towards each other as $\alpha$ approaches
$\alpha_\star$ in \cref{fig:5a} reflects the same. When
$\alpha=\alpha_\star$, we see in \cref{fig:5b} that a minimum of the
energy annihilates the maximum at $\lambda=\lambda_+$. The resulting
equilibrium is no longer stable and there is no longer an energy
barrier for the arch to overcome to switch to the stable equilibrium
at $\lambda=\lambda_-$ in \cref{fig:5a}. The constraint
${\bf r}(\ell)\cdot{\bf e}_y=\lambda$ imposed by the force ${\rm F}_y$
prevents the arch from executing this transition in problem
\cref{eq:3-4}. The constraint is absent in problem \cref{eq:3-3};
hence its solution spontaneously jumps at $\alpha=\alpha_\star$.

\subsection{Experimental observations}
\label{sec:3-2}
Next, we record observations relating dynamically snapping arches with
experimental realizations of the auxiliary problem that inform the
snap-through model.

\begin{figure}[t]
  \centering
  \subfloat[\label{fig:6a}]{\includegraphics[width=0.4\textwidth]{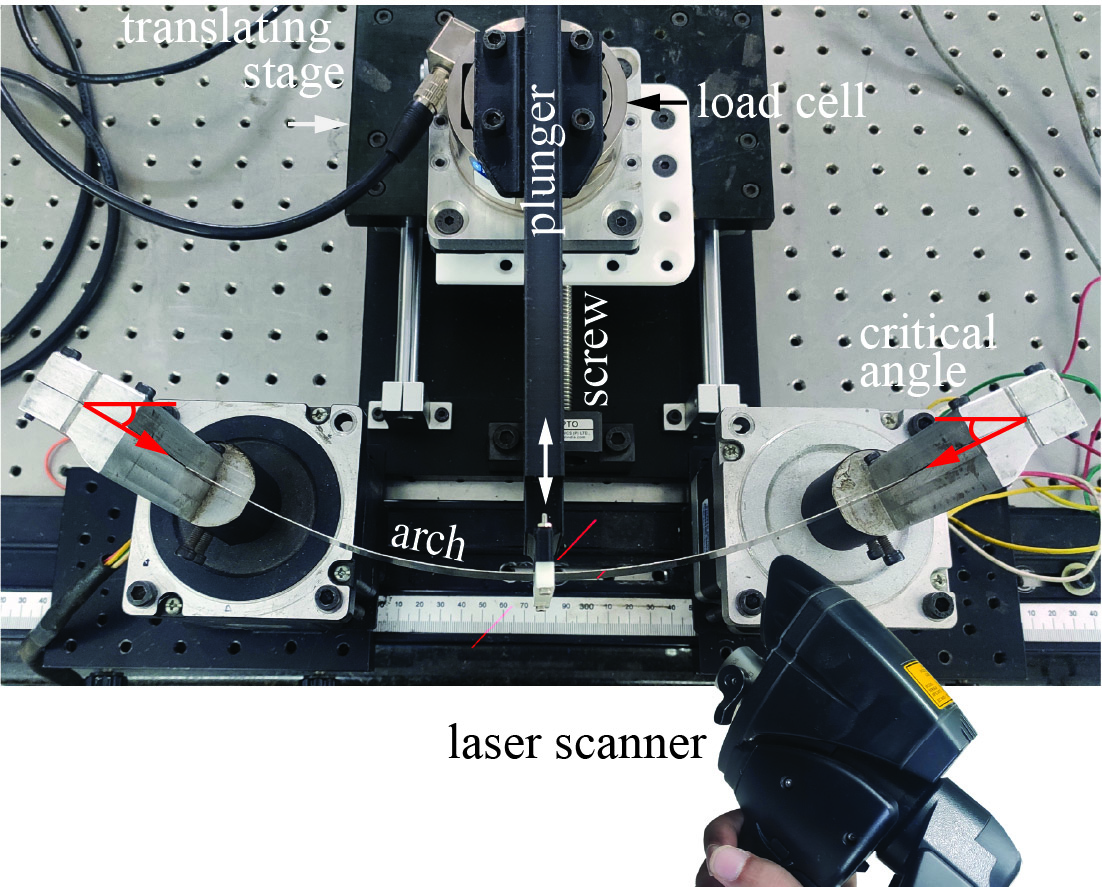}}
  \hfill
  \subfloat[\label{fig:6b}]{\includegraphics[width=0.58\textwidth]{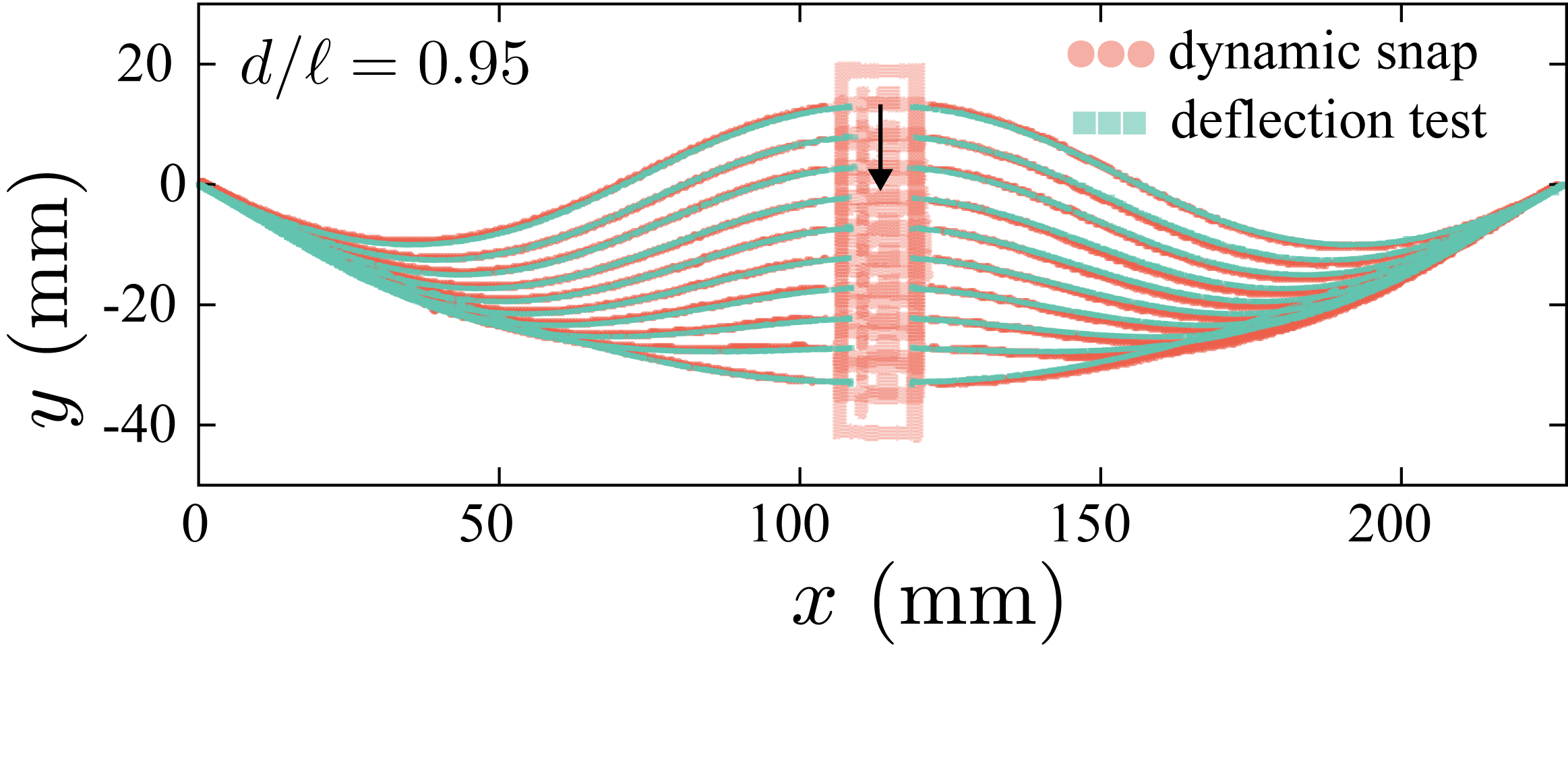}}
  \caption{(a) Experimental realization of the auxiliary problem as a
    quasistatic deflection test. (b) Comparison of arch profiles
    ${\bf r}^{\rm exp}(t_\lambda)$ recorded using a high-speed camera
    during the dynamic snap of the spring steel arch in \cref{fig:1}
    with ${\bf u}^{\rm exp}_\lambda$ measured using a laser scanner in
    the deflection test at coincident arch heights $\lambda$. The
    agreement observed is the basis for the hypothesis underlying the
    proposed model.}
  \label{fig:6}
\end{figure}
\noindent
\paragraph{Experimental realization of the auxiliary problem.} The
auxiliary problem has a simple physical interpretation.  The
inclination $s\mapsto \psi_{\lambda,\alpha}(s)$ is the equilibrium
solution realized in an experiment in which the span is set to $2d$,
the clamps are inclined at angle $\alpha$, and the arch height is
prescribed to be $\lambda$.  In particular, the sequence
$\lambda\mapsto \psi_{\lambda,\alpha}$ realized with fixed span and
clamp orientations while varying $\lambda$ are precisely the solutions
realized in a routine displacement-controlled deflection test
conducted to measure the reaction forces at the clamps as a function
of the arch's height. Of direct significance to us is the special case
of the deflection test with $\alpha$ set to the critical angle. For
convenience, we henceforth refer to the solution
$\psi_{\lambda,\alpha_\star}$ as $\phi_\lambda$ and denote the
corresponding centerline profile by ${\bf u}_\lambda$.

\Cref{fig:6a} shows the experimental setup realizing the auxiliary
problem as a deflection test. The height $\lambda$ is controlled using
a plunger whose one end is rigidly attached to the center of the arch
while the other end is attached to a load cell mounted on a stage that
translates on a linear screw. The figure shows the case $d/\ell=0.95$,
chosen to coincide the compression set for the snapping experiment in
\cref{fig:1}. The end clamps are oriented to the critical angle as
well. Nevertheless, the arch remains at equilibrium due to the
constraint imposed by the plunger, which prevents it from from
snapping. For a discrete set of $\lambda$ ranging between
$\tilde{\lambda}_+\equiv \lambda_+/\ell =0.107$ and
$\tilde{\lambda}_- \equiv \lambda_-/\ell =-0.28$, we record the arch
profiles $\{{\bf u}^{\rm exp}_\lambda\}_\lambda$ realized in the
experiment using a laser scanner having an accuracy of approximately
$80\,\mu{\rm m}$.

\noindent
\paragraph{Dynamic snap and the deflection test.}
The mid point of the arch can be set to travel from
$\lambda=\lambda_+$ to $\lambda_-$ in the deflection test, just as it
does when it spontaneously snaps.  Motivated by this realization, we
compare arch profiles recorded using a high-speed camera during the
snap, with the laser scan data from the deflection tests. The
deflection test being quasistatic rules out juxtaposing measurements
from the two experiments at coincident times. Instead, we do so at
coincident arch heights.  To this end, for
$\lambda\in [\lambda_-,\lambda_+]$, we identify the timestamp
$t=t_\lambda$ from the recording of the dynamic snap at which the arch
height equals $\lambda$. In this way, we determine a sequence of arch
profiles $\{ {\bf r}^{\rm exp}(t_\lambda)\}_\lambda$ indexed by the
arch height in the snap-through experiment. The finite temporal and
spatial resolutions of the camera introduces errors in these
profiles. However, the frame rate of $5000\,{\rm fps}$ and image
resolution of $992\times 356$ used are sufficiently high to prevent
these errors from meaningfully affecting our observations.

\Cref{fig:6b} compares the arch profiles
${\bf r}^{\rm exp}(t_\lambda)$ with ${\bf u}_\lambda^{\rm exp}$ at
representative values of $\lambda$ in the range
$[\lambda_-,\lambda_+]$ for the case $d/\ell=0.95$. Both experiments
were conducted using the same spring steel arch. We observe good
agreement between the two measurements. This is unexpected, even
surprising. The arch snaps dynamically in one experiment but remains
at quasistatic equilibrium in the other.  The mid point of the arch
slides freely along the rail during the snap, but is geometrically
constrained in the deflection test. The profile
${\bf u}^{\rm exp}_\lambda$ is one that minimizes the elastic energy
when the height is set to $\lambda$; the profile
${\bf r}^{\rm exp}(t_\lambda)$, on the other hand, is not even an
equilibrium state. These differences notwithstanding, the figure shows
that arch profiles realized during the dynamic snap and in the
deflection test are well synchronized by the height parameter
$\lambda$. Analogous experiments conducted with parameters
$(d,\ell)=(117,120), (146.5,150)$ and $(147.5,150)\,{\rm mm}$ all
revealed similarly good agreement.  This observation motivates our
central hypothesis:\\[5pt]
\noindent
\begin{minipage}{\linewidth}
\setlength{\fboxsep}{6pt}
\color{black}
\vrule width 2pt
\hspace{6pt}
\begin{minipage}{0.93\linewidth}
  \textbf{Hypothesis:} \emph{The symmetrically snapping arch
    approximately follows a sequence of elastic energy minimizing
    profiles parameterized by its instantaneous height.}
\end{minipage}
\end{minipage}\\

\begin{figure}[t]
  \centering
  \includegraphics[width=\textwidth]{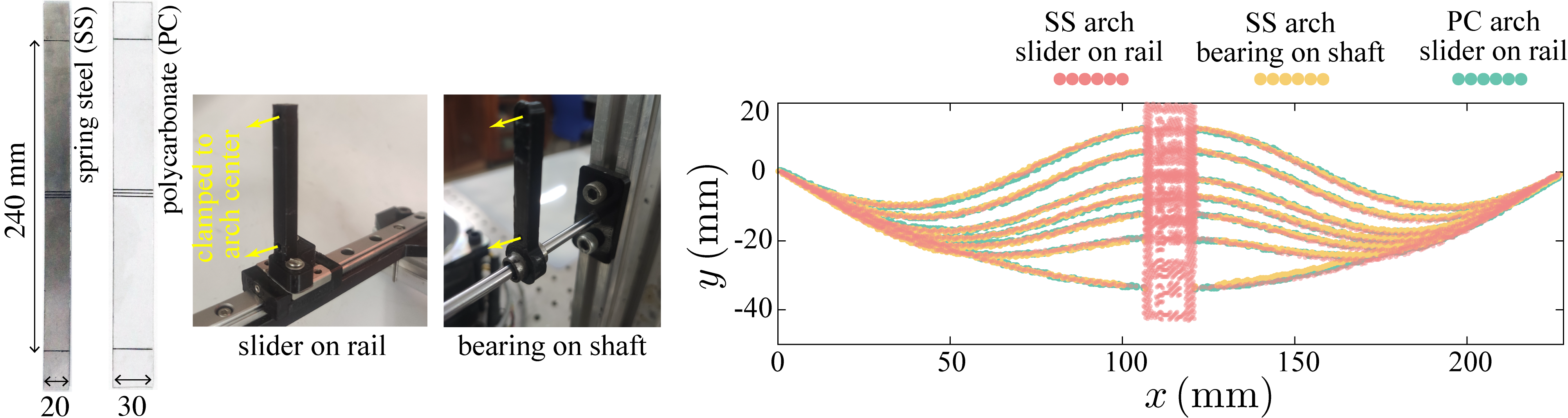}
  \caption{Examining implications of our hypothesis. Replacing the
    spring steel arch with a polycarbonate arch, or the slider block
    attached to the spring steel arch's center with a lighter
    cylindrical bearing does not appear to alter the sequence of
    height-synchronized profiles recorded in the snap-through
    experiments.}
  \label{fig:7}
\end{figure}

Besides the camera's frame rate and image resolution, a couple of other
factors influence the snap-through measurements. As seen in
\cref{fig:1}, the experiment requires attaching a slider block to the
center of the arch to enforce symmetry.  The arch and the slider weigh
$29\,{\rm g}$ and $13\,{\rm g}$, respectively. It is possible,
therefore, that the inertia of the slider and/or the friction between
the rail and the slider bias the agreement observed in
\cref{fig:6b}. This question is best resolved through idealized
dynamic finite element simulations; we do this subsequently. We
investigate the question experimentally by devising an alternate setup
in which we replace the slider-rail assembly with a cylindrical
bearing sliding on a ground shaft. The bearing weighs $5\,{\rm g}$,
less than half of that of the slider block. The friction in the
bearing-shaft assembly is also rated to be much lower than in the
slider-rail arrangement.  \Cref{fig:7} compares measurements recorded
using the two arrangements; the good agreement observed there suggests
that the details of the symmetry-enforcing arrangement in our
snap-through experiments do not influence our hypothesis.

The remark above should not be misconstrued as claiming that the
dynamics of the snapping arch is unaffected by the mass attached to
its center.  The arch is effectively heavier with the slider, and
therefore, we expect it to snap quicker with the lighter bearing
attached. This is indeed what we observe from the camera recordings as
well. The agreement between arch profiles with different masses
attached at the center observed in \cref{fig:7} is despite the arch
snapping at different speeds; synchronizing the two recordings by arch
height rather than time ostensibly factors out the difference in
snapping speeds in the two cases.

\noindent
\paragraph{Material-independence.}
Our hypothesis, claiming an equivalence of arch profiles during
snap-through with those in a deflection test, has a compelling
implication. Consider a pair of arches composed of different
materials, say one of spring steel (SS) and another of polycarbonate
(PC), having identical length $2\ell$ and compressed to the same span
$2d$. Based on our hypothesis, we expect the snap-through of the SS
arch to be described well by the deflection test conducted on it, and
similarly for the PC arch. However, since deflection tests are
displacement-controlled, the profiles realized with the SS and PC
arches are nominally identical. Therefore, our hypothesis predicts
that the SS and PC arches should follow an identical sequence of
profiles during their snap-through.  This realization may appear
perplexing at first, since we expect the material composition to
influence the dynamics of the snap.  Realizing this thought experiment
does show that the SS arch snaps much faster than the PC arch. Yet,
this observation does not contradict our hypothesis, which claims the
snap-through solutions for the SS and PC arches to be synchronized by
height, not in time. Indeed, comparing profiles from high-speed camera
measurements of the snap-through of SS and PC arches at corresponding
heights in \cref{fig:7} validates the material-independent nature of
our hypothesis.

\noindent
\paragraph{Monotonic time history of arch height during snap.} Our
hypothesis relies on synchronizing snap-through solutions with the
energy minimizing solutions of the auxiliary problem in the parameter
$\lambda$. This demands that the evolution of the arch height
$t\mapsto \lambda(t)$ during the snap be an injective function. In
fact, we implicitly assumed this to be the case when indexing dynamic
profiles by $\lambda$ in place of $t$ in the comparisons shown in
\cref{fig:6,fig:7}. \Cref{fig:9b} included later in \cref{sec:4},
tracks the location of the center of the marker pasted on the slider
block in the snap-through experiments with the spring-steel arch. The
plot shows that the arch height does indeed evolve monotonically.

\subsection{Snap-through model}
We now formulate a model for the snap-through. For a given span $2d$,
the critical angle $\alpha_\star$ follows from identifying the turning
point in problem \cref{eq:3-3}. Denote the corresponding pre- and
post-snap arch heights by $\lambda_+$ and $\lambda_-$, respectively.
Without loss of generality, we assume $\lambda_+>\lambda_-$, as has
been the case in our discussions throughout this section. The solution
$s\mapsto \phi_\lambda(s)$ of the auxiliary problem \cref{eq:3-4} with
the clamp angle set to $\alpha=\alpha_\star$ satisfies
\begin{align}
  \begin{cases}
    &\phi_\lambda''+{\rm F}_\lambda\sin\phi_\lambda-{\rm G}_\lambda\cos\phi_\lambda = 0,\\
    &\phi_\lambda(0)=-\alpha_\star, ~\phi_\lambda(\ell)=0, ~\text{and} \\
    &\int_0^\ell(\cos\phi_\lambda,\sin\phi_\lambda)\,ds = (d,\lambda),
  \end{cases} \label{eq:3-6}
\end{align}
where the forces ${\rm F}_\lambda$ and ${\rm G}_\lambda$ are
determined as part of the solution of \cref{eq:3-6} to satisfy the
position constraint specified on the midpoint of the arch. We also
require the \emph{sensitivity}
$\beta_\lambda\equiv d\phi_\lambda/d\lambda$ of the auxiliary problem
to the arch height, determined as the solution of the system
\begin{align}
  \begin{cases}
    &\beta_\lambda'' + ({\rm F}_\lambda\cos\phi_\lambda+{\rm
      G}_\lambda\sin\phi_\lambda)\beta_\lambda + {\rm F}_\lambda'\sin\phi_\lambda-{\rm
      G}_\lambda'\cos\phi_\lambda = 0, \\
    &\beta_\lambda(0) = \beta_\lambda(\ell) = 0, ~\text{and}\\
    &\int_0^\ell(-\sin\phi_\lambda,\cos\phi_\lambda)\beta_\lambda\,ds =
    (0,1).
  \end{cases}
      \label{eq:3-7}
\end{align}
\Cref{eq:3-7} follows from differentiating \cref{eq:3-6} with respect
to $\lambda$.  The parameters ${\rm F}_\lambda'$ and
${\rm G}_\lambda'$ are the sensitivities of the forces
${\rm F}_\lambda$ and ${\rm G}_\lambda$, respectively, and serve to
impose the integral constraint noted in \cref{eq:3-7}. They are
determined alongside $\beta_\lambda$ as part of the solution of
\cref{eq:3-7}.  \Cref{eq:3-7} is a linear system for
$s\mapsto \beta_\lambda(s)$ and
$({\rm F}_\lambda', {\rm G}_\lambda')$. Since it is defined in terms
of $ \phi_\lambda, {\rm F}_\lambda$ and ${\rm G}_\lambda$, problems
\cref{eq:3-6,eq:3-7} are solved in that order.  Henceforth, we assume
solutions $\lambda\mapsto \{\phi_\lambda,\beta_\lambda\}$ of
\cref{eq:3-6,eq:3-7} to be available for each
$\lambda\in [\lambda_-,\lambda_+]$.

For $\lambda_-\leq \lambda \leq \lambda_+$, we postulate
\begin{align}
  \Theta(s,\lambda) \equiv \phi_\lambda(s) ~~ \text{and} ~~
  {\bf R}(s,\lambda) \equiv
  \int_{0}^s(\cos\phi_\lambda(\sigma), \sin\phi_\lambda(\sigma))\,d\sigma \label{eq:3-8}
\end{align}
to be the inclination and the centerline of the snapping elastica, and
the speed of the midpoint of the arch when it is at height $\lambda$
to be
\begin{align}
  \dot{\lambda} = -\sqrt{\frac{{\rm E}[\phi_{\lambda_+}]-{\rm
  E}[\phi_\lambda]}{{\rm M}(\lambda)}},
  ~~\text{where}~\begin{cases}
    {\rm M}(\lambda) = \rho\int_{s=0}^\ell\|{\bf
      m}_\lambda(s)\|^2\,ds, \\
    {\bf m}_\lambda(s) =
    \int_{\sigma=0}^s(-\sin\phi_\lambda,\cos\phi_\lambda)\beta_\lambda\,d\sigma,
  \end{cases}
  \label{eq:3-9}
\end{align}
$\rho$ is the uniform mass density per unit length, and
${\rm E}[\cdot]$ is the elastic energy functional given by
\cref{eq:3-5}. The rationale for \cref{eq:3-8} is provided by the
hypothesis from \cref{sec:3-2}; it faithfully replicates the ansatz
that profiles of a snapping arch are synchronized with solutions of
the auxiliary problem by the height parameter $\lambda$. The transient
nature of the snap is determined by
\cref{eq:3-9}. 
All expressions on the right handside of \cref{eq:3-9} depend only on
solutions $\phi_\lambda$ and $\beta_\lambda$, and are hence
computable. The negative sign in $\dot{\lambda}$ is a consequence of
the sign convention adopted for $\lambda$, whence
$\lambda_+>\lambda_-$.

\noindent
\paragraph{Dynamics from energy conservation.} The rationale for
\cref{eq:3-9} is energy conservation. Postulating that the sum of the
elastic and kinetic energies remains unchanged during the snap, we
have
\begin{align}
  \underbrace{{\rm E}[\Theta]}_{\rm elastic} +
  \underbrace{{\rm KE}[{\bf R};{\bf v}]}_{\rm kinetic} = {\rm E}_0
  ~(\text{constant}), \label{eq:3-10}
\end{align}
where $(s,\lambda)\mapsto {\bf v}(s,\lambda)$ is the velocity field of
the arch. Since $\Theta=\phi_\lambda$, the elastic component of the
energy in \cref{eq:3-10} equals ${\rm E}[\phi_\lambda]$.  Evaluating
\cref{eq:3-10} at the beginning of the snap when $\lambda=\lambda_+$
and the kinetic energy is zero, we conclude that
${\rm E}_0={\rm E}[\phi_{\lambda_+}]$. It remains to compute the
kinetic energy
\begin{align}
  {\rm KE}[{\bf R};{\bf v}] &\equiv \int_{s=0}^\ell\rho\|{\bf
                              v}(s,\lambda)\|^2\,ds, \label{eq:3-11}
\end{align}
where we have ignored the contributions from the arch's rotary
inertia. To evaluate the velocity, we differentiate ${\bf R}$ in
\cref{eq:3-8} using the chain rule to get
\begin{align}
  {\bf v}(s,\lambda)
  &= \frac{d}{dt}{\bf R}(s,\lambda) =
    \dot{\lambda}\frac{\partial}{\partial \lambda}{\bf R}(s,\lambda) 
    = \dot{\lambda}
    \int_{\sigma=0}^s (-\sin\phi_\lambda,
    \cos\phi_\lambda)\,\frac{d\phi_\lambda}{d\lambda}\,d\sigma
    = \dot{\lambda}{\bf m}_\lambda(s). \label{eq:3-12}
\end{align}
Notice that the sensitivity of the auxiliary problem naturally appears
when computing the velocity of the arch. \Cref{eq:3-12} also suggests
interpreting ${\bf m}_\lambda(s)$ as a ``mobility vector'' relating
velocities along the arch to that at its center. Noting \cref{eq:3-12}
in \cref{eq:3-11}, we get
\begin{align}
  {\rm KE}[{\bf R},{\bf v}] = \dot{\lambda}^2\rho \int_{s=0}^\ell
  \|{\bf m}_\lambda(s)\|^2\,ds = \dot{\lambda}^2 {\rm M}(\lambda). \label{eq:3-13}
\end{align}
Combining \cref{eq:3-13} with \cref{eq:3-10}, and using
${\rm E}[\Theta]={\rm E}[\phi_\lambda]$ and
${\rm E}_0={\rm E}[\phi_{\lambda_+}]$ yields
\begin{align}
  {\rm E}[\phi_{\lambda_+}] = {\rm E}[\phi_\lambda] +
  \dot{\lambda}^2 {\rm M}(\lambda)~~\Rightarrow~~\dot{\lambda}^2 =
  \frac{{\rm E}[\phi_{\lambda_+}] -{\rm
  E}[\phi_\lambda]}{{\rm M}(\lambda)}. \notag
\end{align}
\Cref{eq:3-9} then follows from using the negative square-root in the
expression above.

\noindent
\paragraph{Arch states.} \Cref{eq:3-8,eq:3-9} define all aspects of
the arch's snap-through. First, the temporal evolution of the midpoint
is the solution of the initial value problem:
\begin{align}
  \frac{d\lambda}{dt} =  -\sqrt{\frac{{\rm E}[\phi_{\lambda_+}]-{\rm
  E}[\phi_\lambda]}{{\rm M}(\lambda)}}~~\text{with initial condition}~~\lambda(t=t_0) =
  \lambda_+,\label{eq:3-14}
\end{align}
where $t_0$ is the (arbitrary) time stamp at the onset of the snap
when $\lambda=\lambda_+$. The inverse map $\lambda\mapsto t(\lambda)$,
parameterizing time by arch height, is more readily computable from
\cref{eq:3-14}:
\begin{align}
  \lambda\mapsto t(\lambda) = t_0 +
  \int_{\xi=\lambda}^{\lambda_+}\sqrt{ \frac{{\rm M}(\xi)}{{\rm
  E}[\phi_{\lambda_+}]-{\rm E}[\phi_\xi]}}\,d\xi. \label{eq:3-15}
\end{align}

Second, the state $(\lambda,\dot{\lambda})$ of the midpoint fully
determines the dynamic state $({\bf R},{\bf v})$ of the arch
itself. Specifically, ${\bf R}$ is given by \cref{eq:3-8} and
${\bf v}$ by \cref{eq:3-12}. Arch states can alternately be
reparameterized by time instead of $\lambda$. These states, say
$(s,t)\mapsto (\bar{\bf R}(s,t), \bar{\bf v}(s,t))$, follow from the
solution of \cref{eq:3-14} as
$\bar{\bf R}(s,t) = {\bf R}(s,\lambda(t))$ and
$\bar{\bf v}(s,t)={\bf v}(s,\lambda(t))$.

We conclude this section with a few remarks.  \Cref{eq:3-9} is a
reduced order model that formulates the snap-through problem of the
arch as one of computing the state $(\lambda,\dot{\lambda})$ of just
its midpoint. Then, \cref{eq:3-8} fully determines the state
$({\bf R},{\bf v})$ of the arch given $(\lambda, \dot{\lambda})$. The
auxiliary solution $\phi_\lambda$ and the mobility vector
${\bf m}_\lambda$ help accomplish the lifting
$(\lambda,\dot{\lambda})\mapsto ({\bf R}, {\bf v})$.

The model does not introduce any linearization of the elastica's
kinematics; problems \cref{eq:3-6,eq:3-7} fully retain the geometric
nonlinearity of the planar elastica theory. This feature of the model
becomes significant when the arch is subject to reasonably large
compressions. For instance, in \cref{sec:4}, we examine the model's
predictions up to $d/\ell =0.7$. At such compressions, linearized
solutions of \cref{eq:3-6,eq:3-7}, though computed more conveniently,
differ significantly from their nonlinear counterparts.

Finally, the model implicitly assumes that the arch-height evolves
monotonically with time. This assumption permits parameterizing
solutions by $\lambda$ in place of $t$.  \Cref{eq:3-9} shows that the
evolution of the arch height $t\mapsto \lambda(t)$ is injective if
$({\rm E}[\phi_{\lambda_+}]-{\rm E}[\phi_\lambda])/{\rm M}(\lambda)$
remains positive and bounded for $\lambda<\lambda_+$.  We examine this
later using numerical computations. For now, we note that
${\rm E}[\phi_{\lambda_+}]>{\rm E}[\phi_\lambda]$ for
$\lambda_-\leq \lambda<\lambda_+$ follows directly from examining
energies of solutions to the auxiliary problem.  As the integral of
$\|{\bf m}_\lambda\|^2$, ${\rm M}(\lambda)$ is guaranteed to be
non-negative.


\section{Model validation}
\label{sec:4}
We devote this section to studies validating the model, relying
heavily on dynamic FE simulations of snapping arches. The idealized
comparisons with FE simulations avoids uncertainties inherent in the
experiments (dimensional inaccuracies, friction, spatial/temporal
resolution of measurements). Furthermore, the FE simulations help
examine the model's prediction accuracy for quantities not directly
measured in the experiments, such as velocities and energies. The
simulations also enable imposing the symmetry assumed in the problem
without appending intrusive components to the center of the arch. As
discussed previously, the sliders and bearings used in the
snap-through experiments necessarily alter the inertia of the arch and
introduce concerns about the significance of friction forces.

\noindent
\paragraph{Model simulations.}
For a given compression $d/\ell$, we determine the buckled solution by
solving \cref{eq:3-2}, and compute the bifurcation diagram for problem
\cref{eq:3-3} to determine the critical angle $\alpha_\star$, and the
arch heights $\lambda_+$ and $\lambda_-$ at the pre- and post-snap
solutions. Then, for a dense sampling $\{\lambda_i\}_{i=1}^n$ of the
interval $[\lambda_-,\lambda_+]$ satisfying $\lambda_i>\lambda_{i+1}$,
$\lambda_1=\lambda_+$ and $\lambda_n=\lambda_-$, we compute solutions
$\lambda\mapsto (\phi_\lambda,\beta_\lambda)$ of \cref{eq:3-6,eq:3-7}
by treating $\lambda$ as the continuation parameter. In our computer
implementation, we use AUTO-07P \cite{doedel2007auto} to approximate
solutions of all quasistatic problems
\cref{eq:3-2,eq:3-3,eq:3-6,eq:3-7} over a symmetric half of the
arch. Alternatively, shooting methods, finite difference, and finite
element discretizations can be adopted. Elliptic integral solutions
can also be directly leveraged for solving
\cref{eq:3-2,eq:3-3,eq:3-6}.

With solutions $\{\phi_\lambda,\beta_\lambda\}_\lambda$ at hand, a
direct evaluation of \cref{eq:3-8} yields the arch profile at each
$\lambda=\lambda_i$. Next, we evaluate \cref{eq:3-9} to determine the
speeds $\{\dot{\lambda}_i\}_i$ at corresponding arch heights
$\{\lambda_i\}_i$. The elastic and kinetic energies
${\rm E}[\phi_{\lambda_i}]$ and $\dot{\lambda}_i^2{\rm M}(\lambda_i)$
are determined as part of these calculations. Finally, we evaluate
\cref{eq:3-15} and record the time stamps $t(\lambda_i)$.

We highlight that simulating the model only involves solving ODEs for
quasi-static boundary value problems; evaluating the speed
$\lambda\mapsto\dot{\lambda}$ and time parameterization
$\lambda\mapsto t(\lambda)$ are simple integral evaluations. In
particular, simulating the model does not involve discretizing PDEs at
any stage, nor does it introduce considerations of stability of
numerical discretizations.

\noindent
\paragraph{Dynamic finite element simulations.}
We simulate the dynamic snap-through of arches using geometrically
nonlinear beam elements in ABAQUS. Considering the reflection symmetry
about the center, we only simulate one half of the arch. The
simulations then faithfully follow the three-step workflow depicted in
\cref{fig:2}, albeit in the planar setting. Hence, we pre-compress the
arch, causing it to buckle. Then, with the span fixed at $d$, we
quasi-statically rotate the ends to the pre-computed critical angle
$\alpha_\star$ before switching to a dynamic step. In principle, the
arch would snap spontaneously without introducing any additional
perturbations. However, the onset of the snap simulated this way is
difficult to predict, and often results in the arch remaining idle for
long simulation times before snapping rapidly. As a remedy, we rotate
the ends slightly past the critical angle to $\alpha_\star+0.05^\circ$
in the dynamic step to trigger the snap more quickly.

We note that the FE solutions simulated with ABAQUS are not a faithful
discretization of the dynamical elastica theory. The beam element
adopted in the simulations permits axial and shear strains, unlike the
kinematics assumed for the elastica. However, a posteriori inspection
of the simulations confirm that axial and shear strain contributions
to elastic energies remain uniformly small. The arch's deformation
remains bending-dominated during the snap; its slenderness ensures
that the extensional and shear stiffness are much larger than the
bending stiffness

In the remainder of this section, we employ the non-dimensional
parameters
$\tilde{\lambda} = \lambda/\ell, \tilde{\dot{\lambda}} = \sqrt{{\rm
    B}/\rho\ell^2}, \tilde{\rm E}[\cdot] = {\rm E}[\cdot]/({\rm
  B}/\ell)$ and $\tilde{t}=t/\sqrt{\rho\ell^4/{\rm B}}$ when comparing
model predictions with FE simulations. It is straightforward to
express the model \cref{eq:3-8,eq:3-9} in non-dimensional form using
these scalings.

\subsection{Validation at $d/\ell=0.95$}
\label{sec:4-1}
First, we validate the model's prediction at the compression ratio
$d/\ell=0.95$ chosen in the experiments discussed previously in
\cref{sec:3-2}.
\begin{enumerate}
\item \Cref{fig:8a} examines the velocity of the mid point of the arch
  as a function of its height. The results of the model and FE
  simulation agree well. While the model's prediction follows directly
  from integrating \cref{eq:3-9}, we use the time-histories of the
  position and velocity of the mid point from the FE simulation.
  Owing to our sign convention for $\lambda$, the snap proceeds from
  left to right in the plot, i.e., from
  $\tilde{\lambda}=\tilde{\lambda}_+$ to $\tilde{\lambda}_-$. At the
  start of the snap, the speed is zero. Somewhat surprisingly, both
  the model and the FE simulation show that the arch reaches its
  maximum speed at approximately $80\%$ of the travel distance. It
  decelerates thereafter.

\item \Cref{fig:8b} compares the elastic and kinetic energies
  predicted by the model and the FE simulation as a function of the
  arch height. The energy profiles agree well. The elastic energy
  decreases monotonically as the arch snaps, which is accompanied by a
  monotonic rise in the kinetic component.
  The elastic energy predicted by the model in \cref{fig:8b} is
  identical to that computed for the auxiliary problem in
  \cref{fig:5b} at the critical angle. Thus, \cref{fig:8b} shows that
  the snapping arch effectively follows an elastic energy path that is
  {unaware} of its momentum.
  
  \begin{figure}[t]
    \centering
    \subfloat[\label{fig:8a}]{\includegraphics[width=0.32\textwidth]{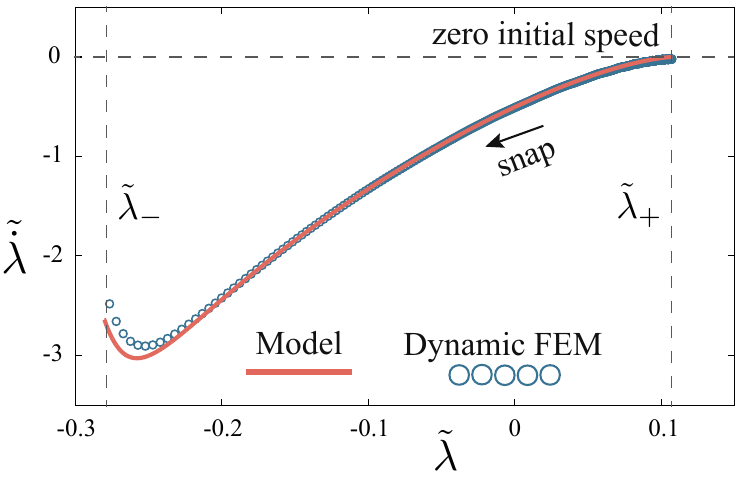}}
    \hfill
    \subfloat[\label{fig:8b}]{\includegraphics[width=0.32\textwidth]{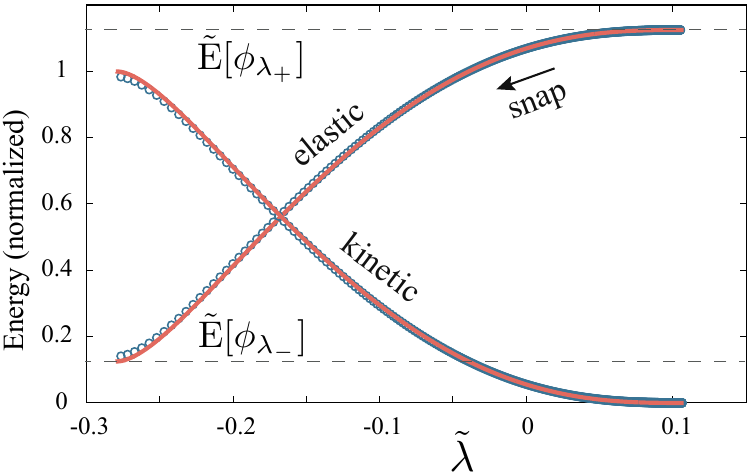}}
    \hfill
    \subfloat[\label{fig:8c}]{\includegraphics[width=0.32\textwidth]{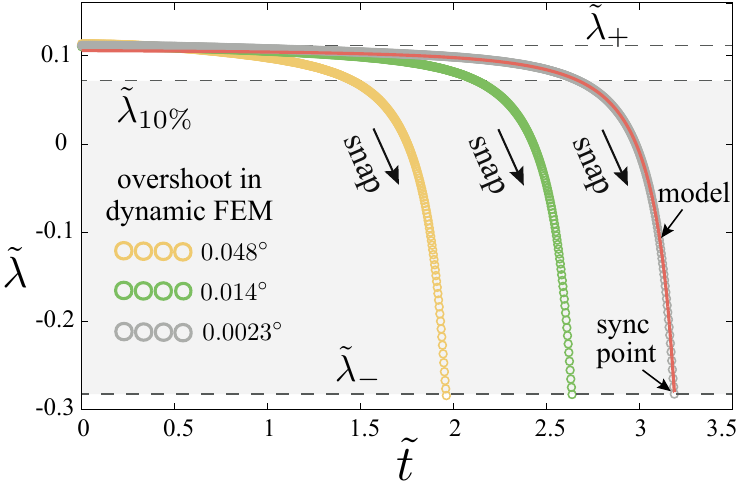}}
    \\
    \subfloat[\label{fig:8d}]{\includegraphics[width=0.32\textwidth]{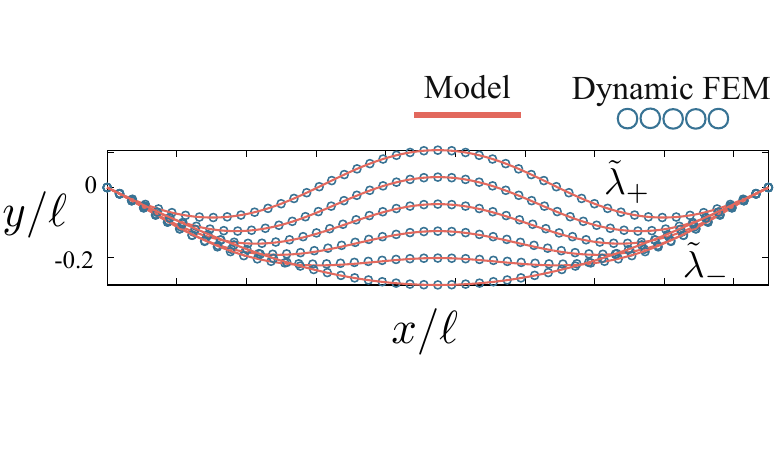}}
    \hfill
    \subfloat[\label{fig:8e}]{\includegraphics[width=0.32\textwidth]{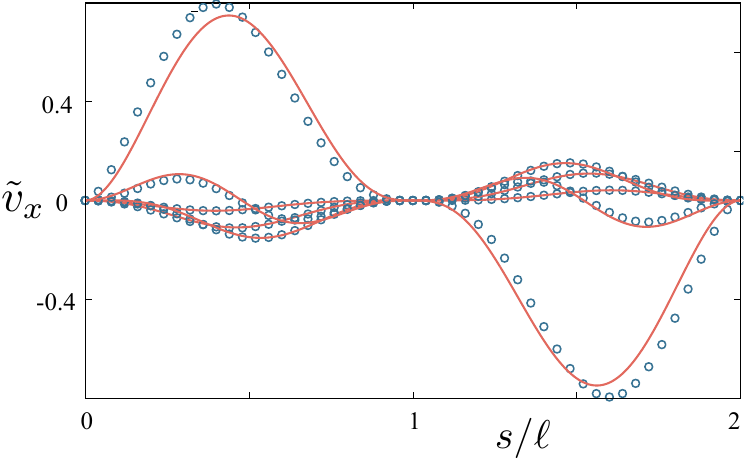}}
    \hfill
    \subfloat[\label{fig:8f}]{\includegraphics[width=0.32\textwidth]{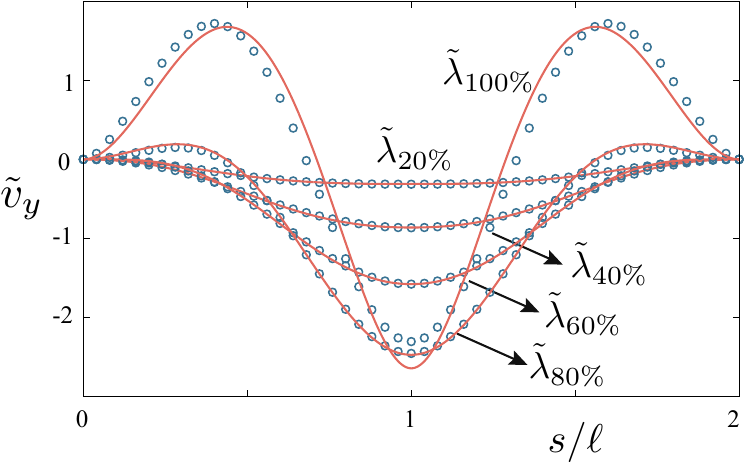}}
  \caption{Validation of the model's predictions with dynamic finite
    element simulations at the compression ratio $d/\ell=0.95$. The
    figure reveals good agreement between the two for mid point
    position (c) and velocity (a), elastic and kinetic energies (b),
    arch profiles (d), and velocity distributions (e,f). All plots use
    non-dimensional parameters; see \cref{sec:4-1} for details.}
  \label{fig:8}
\end{figure}

\Cref{fig:8a,fig:8b} show that the kinetic energy of the arch is not
maximal when the mid point attains its maximum speed, i.e., the
locations of the extrema of the kinetic energy and the speed of the
arch's mid point do not coincide.  The midpoint's speed is maximal
before the end of the snap, while the kinetic energy grows until the
end of the snap.

  Their agreement notwithstanding, we record a few distinctions
  between the model and FE predictions. First, the sum of the elastic
  and kinetic energy components is exactly conserved in the model by
  virtue of \cref{eq:3-10}, but only approximately in the FE
  simulation. Second, the model underestimates the elastic energy
  compared to the FE simulation. This is due to the ansatz in
  \cref{eq:3-8} that the arch adopts the minimal elastic energy
  configuration for a given arch height. As a consequence, the model
  overestimates the kinetic energy.  In particular, the model predicts
  an arch with higher kinetic energy at the end of the snap than the
  FE simulation. The difference, though small, is more apparent
  towards the end of the snap in the plot. Third, the model and the FE
  simulation do not predict the same arch configuration at the end of
  the snap. The model's prediction coincides with the stable
  equilibrium configuration. This is not necessarily the case in the
  FE simulation. However, after sufficiently long times and in the
  presence of energy dissipating mechanisms (either physical or
  numerical), the arch eventually stops vibrating after snapping and
  settles to the same equilibrium configuration in the FE simulation
  as well.

\item Next, \cref{fig:8c} examines the time-history predicted for the
  mid-point of the arch. Before comparison the model with FE
  simulations, we note that the time history of the snap is highly
  sensitive to perturbations \cite{Gomez2016, Radisson2023-2}.  This
  is the case not only in experiments, but in simulations as well. The
  arch can linger in its unstable configuration for an indeterminate
  time before snapping rapidly. This was the primary reason for
  comparing velocities and energies parameterized by arch height,
  rather than time, in \cref{fig:8a,fig:8b}.

  As mentioned previously, we trigger the snap in the FE simulations
  by rotating the ends slightly past the critical angle. \Cref{fig:8c}
  reveals the high sensitivity of the time history of the snap to this
  ``overstep'' in angle; a larger perturbation triggers the snap more
  quickly. Furthermore, the differences between the three FE
  simulations are restricted to a small initial fraction of the travel
  distance. All three FE simulations agree well over approximately
  $90\%$ of the travel distance and effectively only differ by a
  time-translation over this range.

  Our model does not escape the uncertainty highlighted above in the
  FE simulations. This is because the speed $\dot{\lambda}$ increases
  gradually from zero in \cref{eq:3-9}.  Hence, the duration of
  approximately the first $10\%$ of the travel is highly sensitive to
  the time step chosen to integrate \cref{eq:3-9}.

  With these observations in mind, we synchronize the model's
  prediction with the FE simulation at the end point of the snap,
  rather than at the start. Then, the model and the FE simulations
  agree well over $90\%$ of the travel distance; the plot shows an
  overlay of the model with one of the FE simulations to illustrate
  the match.

\item The model furnishes not just the state $(\lambda,\dot{\lambda})$
  of the mid point but of the arch itself. \Cref{fig:8d} compares the
  arch profiles predicted by the model with those from the FE
  simulations at corresponding arch heights, in increments of $20\%$
  of the midpoint's travel. The good agreement of profiles in the
  figure corroborates the experimental observations in \cref{fig:6b}
  comparing measurement from the snap-through and deflection
  tests. \Cref{fig:8d} confirms that that the agreement in
  \cref{fig:6b}, which motivated our main hypothesis, was not an
  artefact of the additional inertia or frictional forces introduced
  by the symmetry-enforcing constraints, but is a feature of the
  snap-through phenomenon. \Cref{fig:8e,fig:8d} compare the horizontal
  and vertical components of the velocity along the arc length of the
  arch as the snap progresses. Just as we observed for the mid-point's
  velocity in \cref{fig:8a}, the model's prediction agrees well with
  the FE simulation for approximately $80\%$ of the travel distance;
  small discrepancies emerge thereafter.

  \begin{figure}[t]
    \centering
    \subfloat[\label{fig:9a}]{\includegraphics[width=0.5\textwidth]{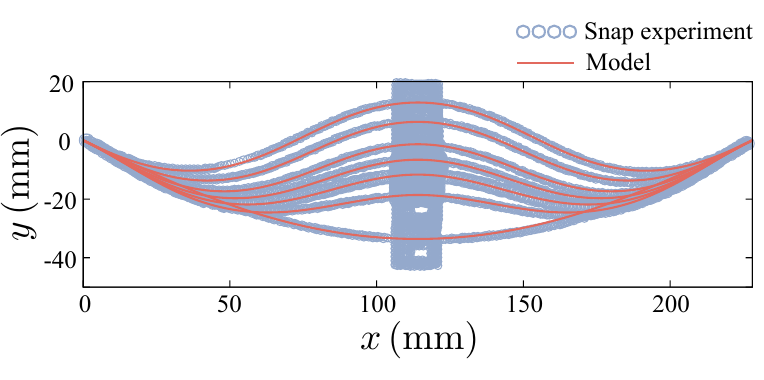}}
    \hfill
    \subfloat[\label{fig:9b}]{\includegraphics[width=0.45\textwidth]{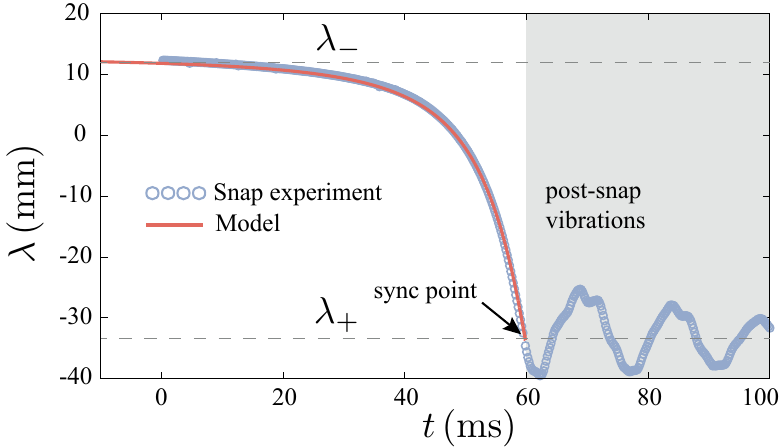}}
    \caption{Validation of the model's predictions by comparison with
      measurements from the dynamic snap-through experiment for the
      compression ratio $d/\ell=0.95$. The time history plotted in (b)
      accounts for the inertia of the slider attached to the center of
      the arch in the experiment.}
    \label{fig:9}
  \end{figure}

\item Complementing the comparisons of the model with the FE
  simulations discussed thus far, \cref{fig:9} validates the model's
  predictions with measurements of the mid-point’s position and arch
  profiles in a snap-through experiment using a spring steel
  arch. \Cref{eq:3-9} does not account for the mass of the slider
  attached to the center of the arch in the experiment. Doing so only
  requires a minor modification to include the kinetic energy of the
  attachment. Setting the mass of the slider to be $4m_0$, we add its
  contribution $m_0\dot{\lambda}^2$ to the kinetic energy for the
  symmetric half of the assembly. The revised expression for the arch
  speed in \cref{eq:3-9} thus requires replacing ${\rm M}(\lambda)$
  with ${\rm M}(\lambda)+m_0$. Attaching a larger mass to the center
  of the arch slows down the snap, as expected. However, \cref{eq:3-8}
  defining arch profiles remains unaffected by the added mass,
  consistent with our observations in \cref{fig:7} comparing
  experiments with different masses attached to the center.
\end{enumerate}

\subsection{Terminal forces}
\label{sec:4-2}
Our description of the model does not yet define the terminal forces
${\bf F} = {\rm F}_x{\bf e}_x+{\rm F}_y{\bf e}_y$ exerted by the
structure on the clamps during the snap-through. In particular, we do
not claim these to be the constraining forces ${\rm F}_\lambda$ and
${\rm G}_\lambda$ appearing in the definition of the quasistatic
auxiliary problem in \cref{eq:3-6}. Doing so would be unsatisfactory
because we expect the momentum of the arch to contribute significantly
to the force history. Instead, we use the profiles and velocities
predicted by the model in a spatially-integrated version of the
statement of momentum balance. Recall that the balance of linear
momentum is given in terms of the stress resultant
${\bf f}(s,t) = f_x(s,t){\bf e}_x + f_y(s,t){\bf e}_y$ as:
\begin{align}
  \rho\frac{d^2\bar{\bf R}(s,t)}{dt^2} = \frac{\partial{\bf
  f}(s,t)}{\partial s}. \label{eq:4-1}
\end{align}
Integrating the component of \cref{eq:4-1} along ${\bf e}_x$ over the
full extent of the arch yields
\begin{align}
  {\rm F}_x(t)= \frac{1}{2}\int_{s=0}^{2\ell}\frac{\partial f_x(s,t)}{\partial s}\,ds =
  \frac{\rho}{2}\int_{s=0}^{2\ell}  \frac{d^2\bar{\rm R}_x(s,t)}{dt^2}\,ds,
\label{eq:4-2a}
\end{align}
where we have used ${\rm F}_x(t) = f_x(0,t) = -f_x(2\ell,t)$ due to
the reflection symmetry about the center. Similarly, integrating the
component of \cref{eq:4-1} along ${\bf e}_y$, we get
\begin{align}
  {\rm F}_y(t) = \int_{s=0}^\ell \frac{\partial f_y(s,t)}{\partial
  s}\,ds =  \rho\int_{s=0}^\ell \frac{d^2\bar{\rm R}_y(s,t)}{dt^2}\,ds \label{eq:4-2b}
\end{align}
by using ${\rm F}_y(t) = f_y(0,t)$ and the symmetry condition
$f_y(\ell,t) =0$.  Evaluating the terminal forces this way, by
integrating the inertial forces, requires the acceleration
$d^2\bar{\bf R}/dt^2$. Although slightly tedious, the calculation is
straightforward. From \cref{eq:3-12}, we have
\begin{align}
  \frac{d^2\bar{\bf R}(s,t)}{dt^2} = \frac{d\bar{\bf v}(s,t)}{dt} =
  \dot{\lambda}\frac{\partial{\bf v}(s,\lambda)}{\partial\lambda} = \dot{\lambda}
  \frac{\partial}{\partial\lambda}\left(\dot{\lambda}{\bf m}_{\lambda}(s)\right) =
  \dot{\lambda}\left( \frac{d\dot\lambda}{d\lambda} {\bf m}_\lambda + \dot{\lambda}\frac{d{\bf
  m}_\lambda(s)}{d\lambda}\right). \label{eq:4-3}
\end{align}
The term $\dot{\lambda}(d\dot{\lambda}/d\lambda)$ appearing in
\cref{eq:4-3} equals $d\dot\lambda/dt$ and hence represents the
acceleration of the center of the arch. To evaluate it, we use
\cref{eq:3-9} to get
\begin{align}
  &\frac{d\dot{\lambda}}{d\lambda}
    = -\frac{d}{d\lambda}\sqrt{\frac{{\rm
    E}[\phi_{\lambda_+}]-{\rm E}[\phi_\lambda]}{{\rm
    M}(\lambda)}}, \notag
\end{align}
where the derivatives of ${\rm E}[\phi_\lambda]$ and
${\rm M}(\lambda)$ are computed from their definitions in
\cref{eq:3-5,eq:3-9}, as
\begin{align}
  \frac{d{\rm E}[\phi_\lambda]}{d\lambda}
  = 2 {\rm B}\int_{0}^\ell\phi'_\lambda\beta'_\lambda\,ds   \quad
  \text{and} \quad
  \frac{d{\rm M}(\lambda)}{d\lambda}
  = 2\rho\int_0^\ell {\bf  m}_\lambda\cdot \frac{d{\bf
  m}_\lambda}{d\lambda}\,ds. \label{eq:4-4}
\end{align}
\begin{figure}
  \centering
  \subfloat[\label{fig:10a}]{\includegraphics[width=0.4\textwidth]{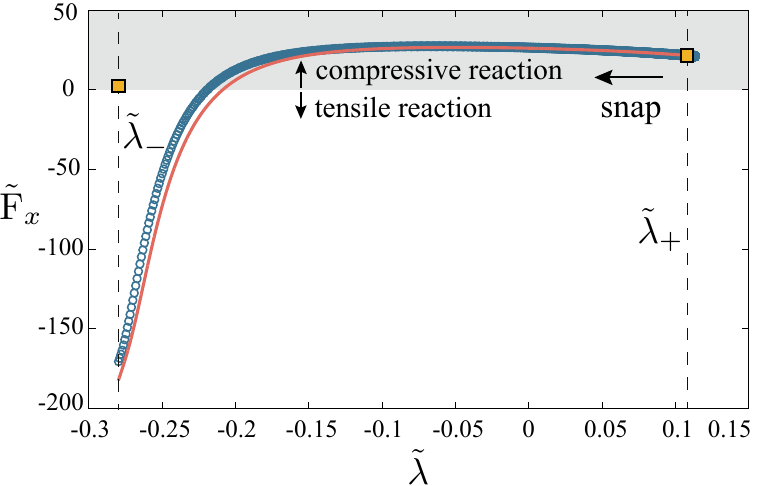}}
  \hspace{0.3in}
  \subfloat[\label{fig:10b}]{\includegraphics[width=0.4\textwidth]{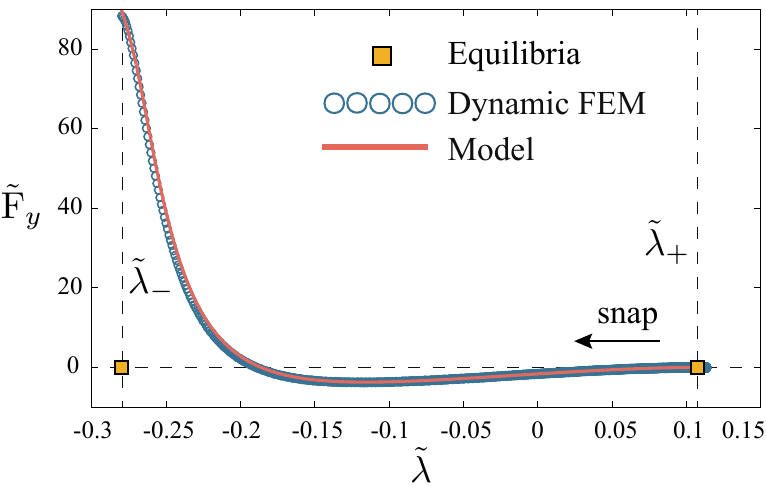}}
  \caption{Comparison of terminal forces predicted by the model using
    \cref{eq:4-2a,eq:4-2b} with those computed using the FE
    simulation. The plots also highlight the significant force
    amplification over the equilibrium reactions, caused by the
    momentum of the arch during the snap.}
  \label{fig:10}
\end{figure}
To compute the sensitivity of the mobility vector required in
\cref{eq:4-3,eq:4-4}, we use \cref{eq:3-9} to get
\begin{align}
  \frac{d{\bf m}_\lambda(s)}{d\lambda}
  &=
    \frac{d}{d\lambda}\int_0^s(-\sin\phi_\lambda,\cos\phi_\lambda)\beta_\lambda\,d\sigma
    \notag \\
  &=
    -\int_0^s(\cos\phi_\lambda,\sin\phi_\lambda)\beta_\lambda\,d\sigma
    +
    \int_0^s(-\sin\phi_\lambda,\cos\phi_\lambda)\frac{d\beta_\lambda}{d\lambda}\,d\sigma, \notag
\end{align}
where $\gamma_\lambda\equiv d\beta_\lambda/d\lambda$ is the solution
of the linear ODE defined by differentiating the system in
\cref{eq:3-7} with respect to $\lambda$, namely,
\begin{align}
  \begin{cases}
    &\gamma_\lambda'' +  ({\rm
      F}_\lambda\cos\phi_\lambda+{\rm
      G}_\lambda\sin\phi_\lambda)\gamma_\lambda  +  ({\rm F}_\lambda''\sin\phi_\lambda-{\rm
      G}_\lambda''\cos\phi_\lambda)  \label{eq:4-5}  \\
    &\qquad = ({\rm
      F}_\lambda\sin\phi_\lambda-{\rm
      G}_\lambda\cos\phi_\lambda)\beta_\lambda^2 
    - 2({\rm F}_\lambda'\cos\phi_\lambda+{\rm
      G}_\lambda'\sin\phi_\lambda)\beta_\lambda,  \\
    &\gamma_\lambda(0) = \gamma_\lambda(\ell) = 0 ~\text{and}~
    \int_0^\ell(-\sin\phi_\lambda,\cos\phi_\lambda)\gamma_\lambda\,ds
    =
    \int_0^\ell(\cos\phi_\lambda,\sin\phi_\lambda)\beta_\lambda^2\,ds. 
  \end{cases} 
\end{align}
In addition to $\gamma_\lambda$, the parameters ${\rm F}_\lambda''$
and ${\rm G}_\lambda''$ in \cref{eq:4-5} are determined as part of
the solution to satisfy the linear integral constraint imposed.

\Cref{fig:10} compares the terminal force histories computed using
\cref{eq:4-2a,eq:4-2b} with those from the FE simulations. We observe
good agreement between the two. The figure additionally indicates the
reaction forces exerted on the clamps by the arch when at its pre- and
post-snap equilibrium states. \Cref{fig:10a} shows that the horizontal
component at both equilibria are positive, implying that the clamp
exerts a compressive force on the arch; the force at the post-snap
equilibrium is smaller in magnitude. However, the dynamic history of
the force component shows that at the end of the snap, the force is
significantly larger in magnitude and is of the opposite sign. The
clamp hence exerts a large tensile force on the arch at the end of the
snap. Similarly, in \cref{fig:10b}, we see a large vertical component
of the force, unlike the zero force at the equilibrium state. These
observations highlight the significant contribution of the momentum of
the arch to the reaction forces, and serve as a reminder against
relying solely on equilibrium calculations in applications harnessing
snap-through instabilities.

\begin{figure}[t]
  \centering
  \subfloat[\label{fig:11a}]{\includegraphics[width=0.32\textwidth]{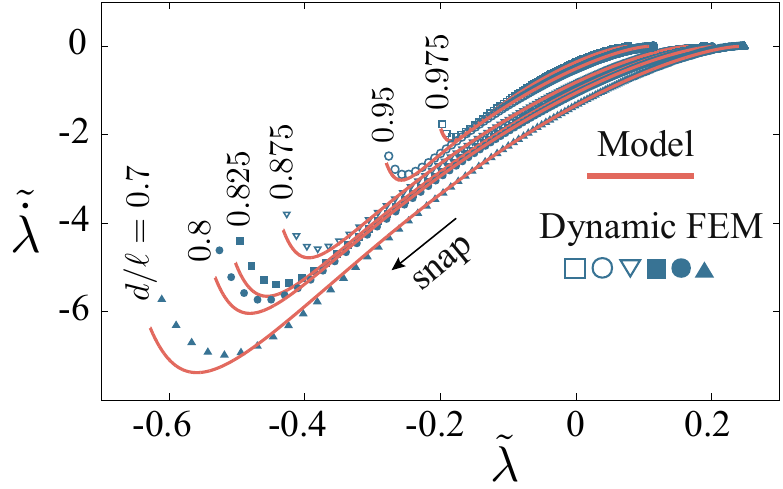}}
  \hfill
  \subfloat[\label{fig:11b}]{\includegraphics[width=0.32\textwidth]{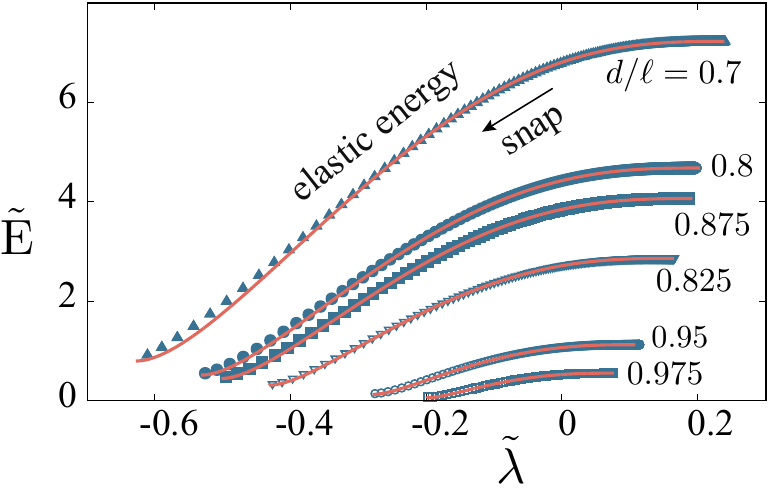}}
  \hfill
  \subfloat[\label{fig:11c}]{\includegraphics[width=0.32\textwidth]{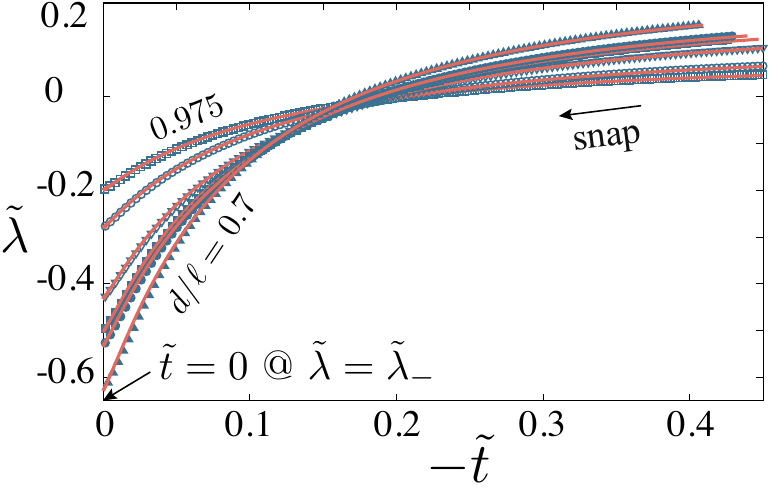}}
  \\
  \subfloat[\label{fig:11d}]{\includegraphics[width=\textwidth]{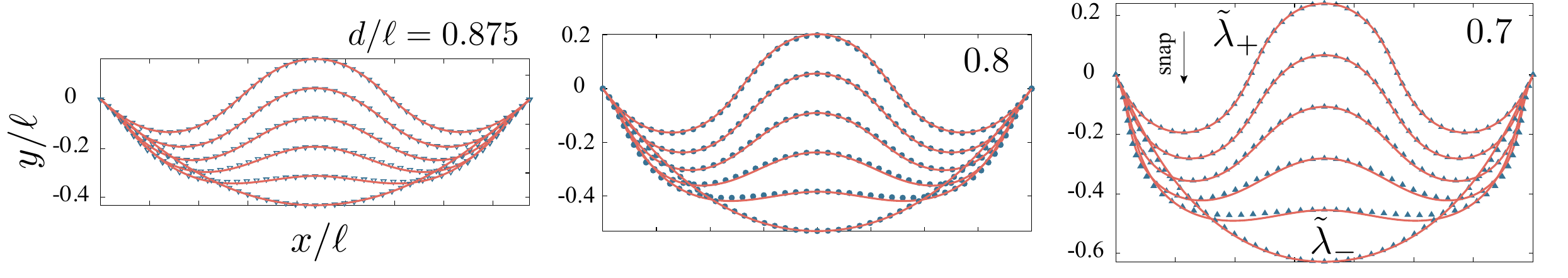}}
  \\
  \subfloat[\label{fig:11e}]{\includegraphics[width=\textwidth]{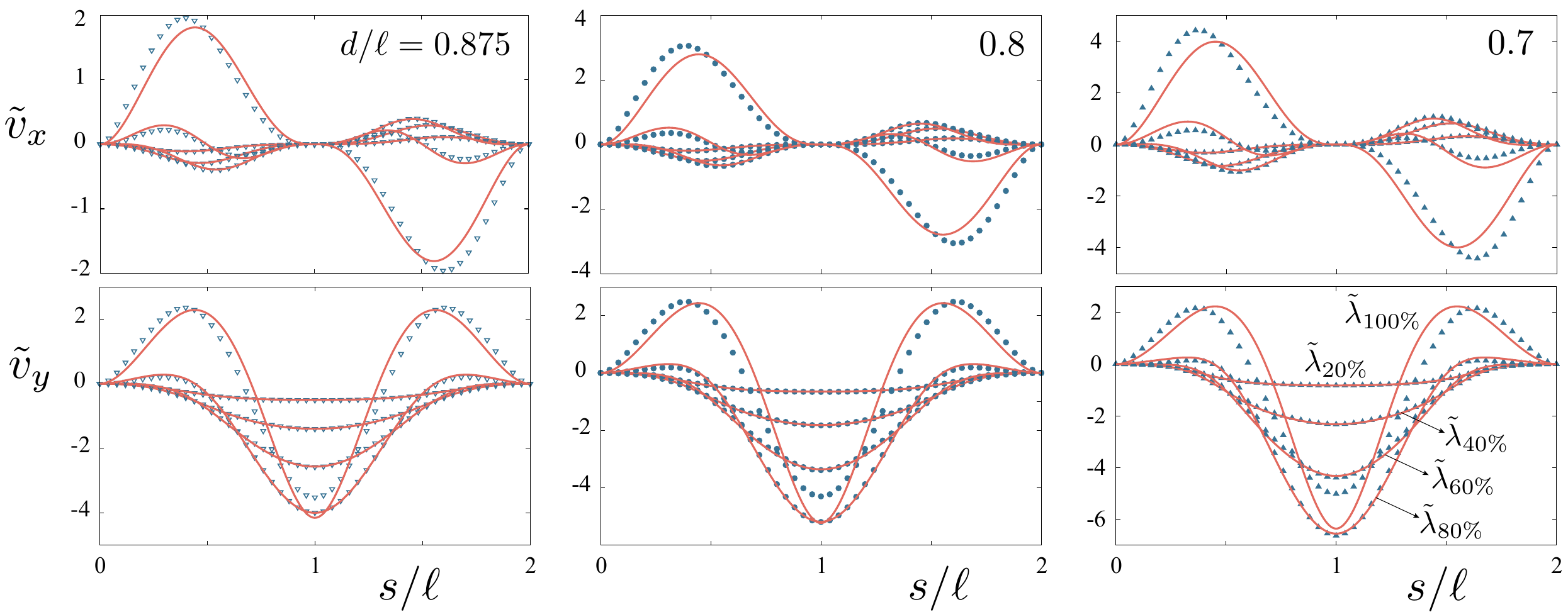}}
  \caption{Comparisons of model predictions with FE simulations over a
    range of compression ratios; see \cref{sec:4-3} for details.}
  \label{fig:11}
\end{figure}

\subsection{Validation with FE simulations at varying $d/\ell$}
\label{sec:4-3}
Next, we validate the model's performance at other compressions using
FE simulations. These simulations help us examine compression ratios
$(d/\ell)$ that are not easily accessible in the experiments.  For
instance, our experimental trials with smaller $d/\ell$ ratios often
resulted in permanent damage in the form of yielding or fracture in
the sample after a few snap-through cycles. While this can be
alleviated by using thinner samples to limit the strain, imaging their
motion is more difficult. Besides, the influence of the
symmetry-imposing arrangement gets exacerbated as well.

\Cref{fig:11} tabulates the performance of the model at compressions
ranging from $d/\ell=0.975$ to $0.7$. The comparisons with FE
simulations largely echo the observations in \cref{fig:9}.
The prediction of the mid-point's velocity shown in \cref{fig:11a}
agrees well with the simulation until the deceleration phase towards
the end of the snap. The elastic energy is predicted well, as revealed
by \cref{fig:11b}, although minor discrepancies are visible towards
the end of the snap for $d/\ell=0.7$. The time history of the
midpoint's motion is plotted in \cref{fig:11c}. For each choice of
$d/\ell$, the synchronization instant $t=0$ for the model and FE
simulation is set to be the end of the snap; hence, the horizontal
axis in the plot is $-\tilde{t}$. For reasons discussed in
\cref{sec:4-1}, the plot does not include the initial $10\%$ of travel
of the arch. \Cref{fig:11d} compares the arch profiles at successive
$20\%$ travel increments. Even at the severe compression ratio of
$d/\ell=0.7$, the quasistatic auxiliary problem's solution agrees well
with the FE simulations.
\Cref{fig:11e} shows that the predicted velocities agree well with the
simulations over about $80\%$ of the travel range, just as observed in
\cref{fig:8e,fig:8f}.

We conclude this section noting that with the exception \cref{fig:9}
comparing model predictions with experimental measurements in absolute
units, all validation studies used non-dimensional parameters and none
required any data fitting.  In fact, we did not even require
assuming/measuring any material property (elastic modulus or mass
density)--- a consequence of the problem studied being
displacement-controlled.
  

\section{Discussion}
\label{sec:5}
We record a few remarks on the model and note observations that
provide additional insights on the snap-through problem studied.

\begin{enumerate}
  
\item The model is phenomenological. The central premise behind it is
  a hypothesis motivated by experimental observations.  The
  comparisons with FE simulations strongly support the model's
  predictive accuracy at significant compressions that are likely
  beyond what may be required in applications requiring the arch to
  execute repeated snap-throughs, such as that related to swimming
  discussed in \cref{sec:2}.

\item The model relies on the quasistatic auxiliary problem to
  construct snap-through solutions.  This should not be misinterpreted
  to mean that the model predicts a quasistatic evolution for the arch
  during the snap. In a sense, the opposite is true--- by postulating
  elastic energy minimizing configurations at each arch height, the
  model overestimates the kinetic energy. The favorable comparisons
  with the dynamic FE simulations shown in \cref{sec:4} should dispel
  any notion of model solutions being quasi-static. Furthermore, the
  snap-through is quick in an absolute sense. For instance, the speed
  of the midpoint of the arch is approximately $6\,{\rm m/s}$ in the
  experiment shown in \cref{fig:9} for the compression ratio
  $d/\ell=0.95$. For the same arch with a compression ratio of $0.7$,
  \cref{fig:11e} shows that the speed of the arch exceeds
  $6\sqrt{{\rm B}/(\rho\ell^2)} \approx 25\,{\rm m/s}$.

\item The hypothesis underlying the model can be interpreted in two
  parts. First, it claims that the snap-through problem can be
  re-parameterized by arch height $\lambda$ in place of time.  Second,
  the arch profile at a given height coincides with energy-minimizing
  configurations. Both are supported by experimental observations in
  \cref{sec:3} and the validation studies in \cref{sec:4}. But neither
  is immediately intuitive.

  \begin{figure}[t]
    \centering
    \subfloat[\label{fig:12a}]{\includegraphics[width=0.52\textwidth]{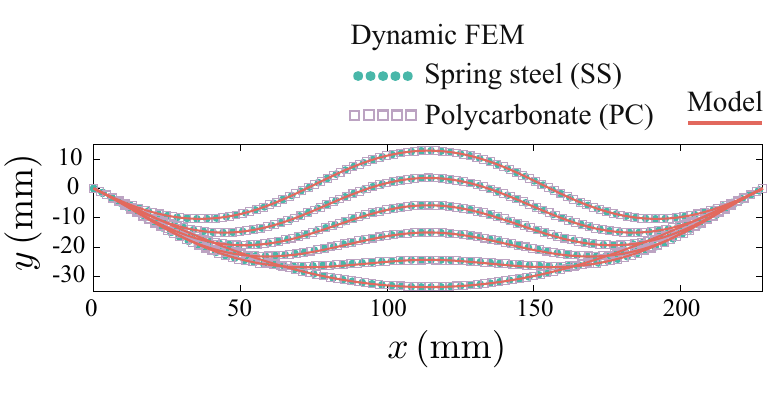}}
    \hfill
    \subfloat[\label{fig:12b}]{\includegraphics[width=0.45\textwidth]{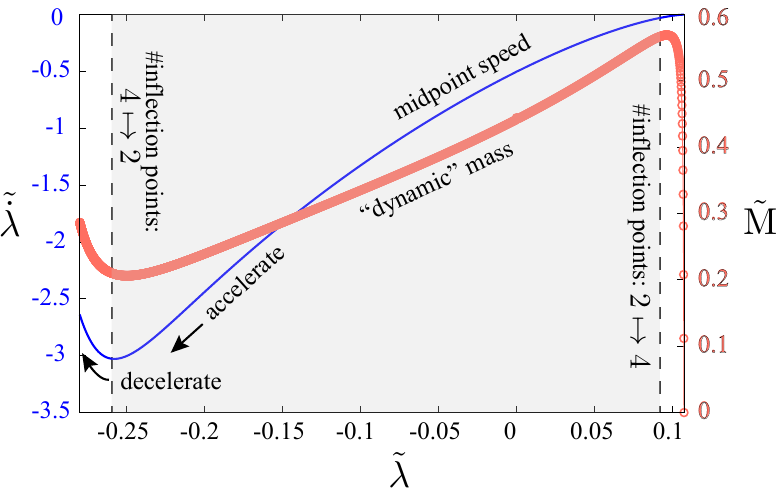}}
    \caption{(a) Comparison of height-synchronized profiles during the
      snap-through of geometrically identical spring steel and
      polycarbonate arches. (b) The function ${\rm M}(\lambda)$
      defined in \cref{eq:3-9} defines an effective mass in the
      model. The plot shows it to be non-constant, and its increase
      towards the end of the snap coincides well with the deceleration
      of the arch and the transition from four to two inflection
      points in the arch's profile.}
    \label{fig:12}
  \end{figure}

\item A corollary of the previous remark is that the model effectively
  delineates the time-independent and transient aspects of the
  snap-through solution.  It does so by reparameterizing solutions as
  $t\mapsto \lambda(t)\mapsto \phi_{\lambda(t)}\mapsto {\bf
    R}(\cdot,\lambda(t))$, rather than, say, by additive superposition
  into spatial and temporal components, or by a separation of
  variables.

  The agreement between profiles realized in snap-through experiments
  with arches composed of different materials noted in \cref{fig:7}
  directly supports the notion that solutions depend on time
  implicitly through the arch height. We corroborate this observation
  in \cref{fig:12a} by repeating the study using dynamic FE
  simulations of arches subject to identical compression but assigned
  material properties (elastic modulus and mass density) corresponding
  to spring steel and polycarbonate, respectively.  Just as our
  hypothesis predicted, the simulations also show that the solutions
  are well-synchronized by arch height. They also agree well with our
  model's prediction, which does not distinguish between the two
  scenarios.
  
\item The ansatz of the arch following elastic energy minimizing
  configurations implies that the model can be interpreted as one that
  \emph{maximizes the elastic energy release rate} during the snap,
  with the rates measured in terms of arch height rather than time. Of
  course, the elastic energy released is converted to kinetic energy
  (rather than being dissipated, as is the case in crack propagation).

\item The dynamic FE simulations reveal small damped oscillations in
  the arch close to the onset of the snap, just as noted in
  \cite{Radisson2023-2}. However, we did not observe such oscillations
  in our experiments, likely due to the influence of the slider-rail
  arrangement and/or the limited spatial resolution of the recording
  camera. Our assumption that the arch's height evolves monotonically
  with time overlooks this aspect of the snap-through.

\item Since the arch is not subject to distributed loads/moments, its
  linear and angular momenta are conserved pointwise.  As a single dof
  description, the model does not (and cannot) impose these. By
  relying on the statement of energy conservation for the entire
  structure to arrive at \cref{eq:3-9}, the model preserves an
  integral of the motion instead.

\item Being a reduced order description, we expect the model to resort
  to some notion of mass lumping to account for the inertia of the
  arch using just the state $(\lambda,\dot{\lambda})$ of the
  midpoint. This is revealed by \cref{eq:3-13}, where the kinetic
  energy of the arch is computed to be
  ${\rm M}(\lambda)\dot{\lambda}^2$. The expression shows
  ${\rm M}(\lambda)$ to be an effective ``dynamic mass'' at the center
  of the arch. \Cref{fig:12b} plots its non-dimensionalized version
  $\tilde{\rm M} = {\rm M}/(\rho\ell)$. The plot reveals that the
  lumped mass is not constant during the snap. Furthermore, the
  location of the minimum of ${\rm M}$ before the end of the snap
  approximately coincides with the extremum of the arch speed
  $\dot{\lambda}$. This suggests that the deceleration observed before
  the end of the snap in \cref{fig:8a,fig:11a} can be attributed to an
  increased effective mass. Curiously, we also find that that the
  locations of these extrema of $\dot{\lambda}$ and ${\rm M}$ agree
  well with the transition from four to two inflection points in the
  arch's profile. This observation is likely a consequence of the
  effective mass in \cref{eq:3-9} being intimately related to the
  shape-sensitivity of the arch's profile to $\lambda$.

\item Hamilton's principle predicts the exact snap-through trajectory
  of the arch as a minimizer the action integral
  ${\rm S}[{\bf r}] = \int_{t_+}^{t_-}{\cal L}({\bf r},\dot{\bf
    r})\,dt$, where as usual, the Lagrangian ${\cal L}$ is the
  difference between the kinetic and elastic energies and $[t_+,t_-]$
  denotes the time duration of the snap. It is then instructive to
  evaluate the action integral over model predictions and compare them
  with those evaluated with the dynamic FE solutions. However, the
  high sensitivity of the snap duration to perturbations renders a
  direct comparison meaningless. Instead, we restrict the evaluation
  of the action integral to the time duration of the last $80\%$ of
  its travel. A caveat of such a comparison is that the start and end
  states of the arch in the model and the FE solutions are slightly
  different. This discrepancy notwithstanding, \cref{fig:13a} compares
  the non-dimensionalized truncated action integral
  $\tilde{\rm S}_{80\%}$ evaluated over the model's prediction and the
  FE simulations for $d/\ell$ ranging from $0.7$ to $0.99$. The good
  agreement observed suggests that the model's predictive accuracy can
  be attributed to it constructing good candidate trajectories that
  closely approximate the minima of the integral over a wide range of
  compression ratios.

  \begin{figure}[t]
    \centering
    \subfloat[\label{fig:13a}]{\includegraphics[width=0.45\textwidth]{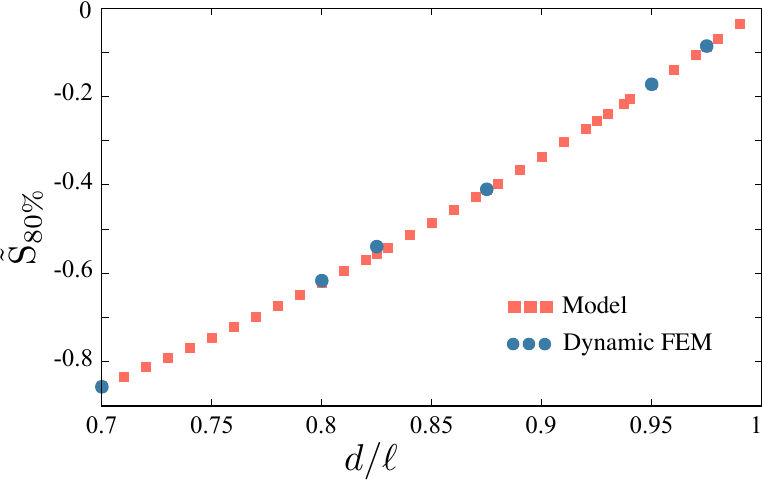}}
    \hspace{0.2in}
    \subfloat[\label{fig:13b}]{\includegraphics[width=0.45\textwidth]{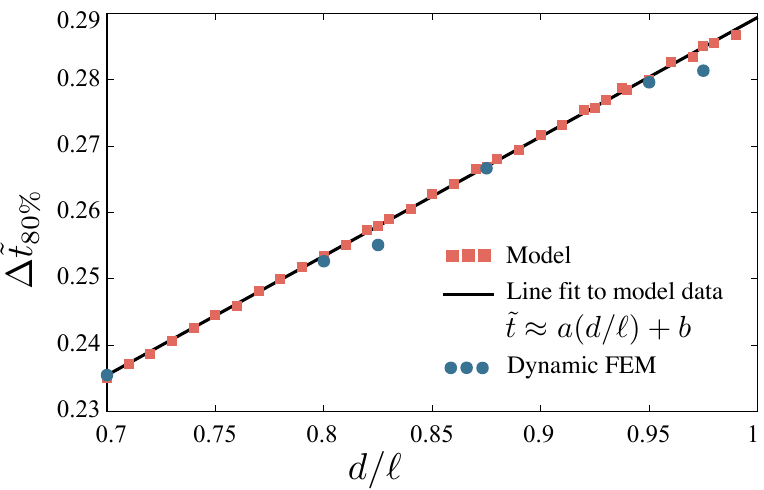}}
    \\
    \subfloat[\label{fig:13c}]{\includegraphics[width=0.45\textwidth]{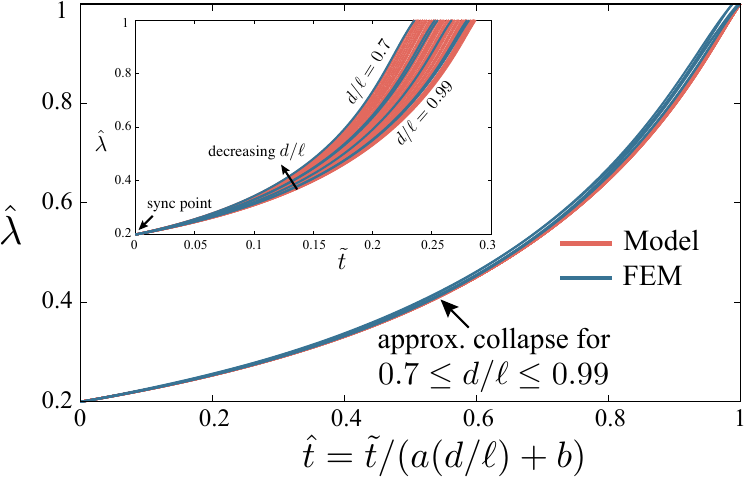}}
    \hspace{0.2in}
    \subfloat[\label{fig:13d}]{\includegraphics[width=0.45\textwidth]{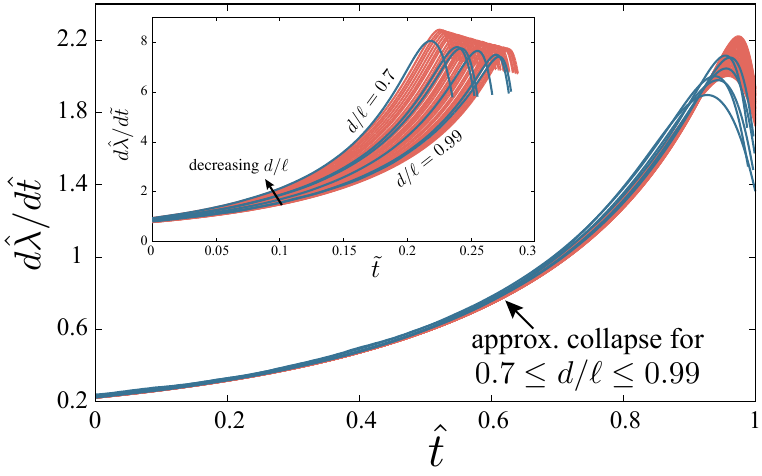}}
    \caption{(a) Examining the truncated action integral computed over
      $80\%$ of the last travel distance suggests that the model
      furnishes good candidate trajectories approximating the
      minimizer of the action integral. (b) The duration for the last
      $80\%$ of the travel during the snap follows an affine trend
      with the compression $d/\ell$. (c,d) Rescaling time by the fit
      found in (b) causes the histories of the arch's midpoint
      position and velocity computed for various compressions to
      approximately collapse.}
    \label{fig:13}
  \end{figure}

\item In the process of evaluating the truncated action, we recorded
  the travel duration $\Delta{t}_{80\%}$ required by the arch to
  traverse the range $\lambda=0.8\lambda_++0.2\lambda_-$ to
  $\lambda_-$ (ignoring the time take for the initial $20\%$ of the
  distance travelled).  \Cref{fig:13b} plots the (non-dimensionalized)
  travel time as a function of the compression ratio $d/\ell$. Besides
  the reasonable agreement observed between the times determined from
  the model and the dynamic FE simulations, the data reveals an affine
  dependence
  \begin{align}
    \Delta \tilde{t}_{80\%} \approx a(d/\ell) + b,
    ~\text{with}~a=0.1798~\text{and}~b=0.1096. \label{eq:5-1}
  \end{align}

  A rather surprising consequence of this affine dependence is shown
  by \cref{fig:13c,fig:13d}. The inset in the former plot shows the
  time-history of the normalized arch height
  $\hat{\lambda} = (\lambda_+-\lambda)/(\lambda_+-\lambda_-)$ for
  various ratios $d/\ell$. The arch height evolution with time is
  steeper at smaller values of $d/\ell$, as expected from arches
  snapping quicker at higher compressions. A similar spread observed
  in the time derivatives plotted in the inset of \cref{fig:13d} shows
  the accelerations to be larger with more compressed arches. However,
  rescaling the non-dimensionalized time parameter $\tilde{t}$ to
  $\hat{t} = \tilde{t}/(a(d/\ell)+b)$ causes the time histories of the
  arch height and its speed to approximately collapse onto a single
  curve.

  The significance of the affine fit in \cref{fig:13b} and the
  collapsed curves in \cref{fig:13c,fig:13d} is that they are
  independent of the dimensions and the material constitution of the
  arch. Hence, they could serve as design aids in choosing the arch’s
  compression level in applications harnessing the instability.

\item We are optimistic that the hypothesis supporting the model can
  be rationalised systematically.  It also remains to be seen if
  aspects of the model and its underlying hypothesis, which we have
  examined for a specific problem, could be helpful in the study of a
  broader class of snap-through instabilities. Preliminary FE
  simulations suggest that the ansatz of the arch following
  energy-minimizing configurations is reasonable in asymmetrically
  snapping planar arches, as well as in the case of the
  three-dimensional snap-through of narrow ribbons, such as those used
  in the swimming application that motivated the model in the first
  place.
\end{enumerate}


\section{Summary}
\label{sec:6}
\begin{figure}
  \centering
  \includegraphics[width=0.6\textwidth]{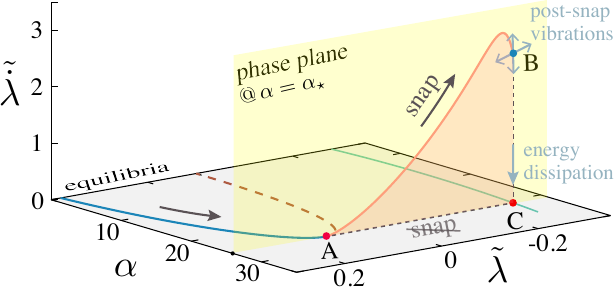}
  \caption{A graphical summary conveying that the model enhances the
    existing view of the snap-through at a fold by predicting the
    dynamics in the phase plane at the critical angle.}
  \label{fig:14}
\end{figure}

We conclude the article with the graphical summary in \cref{fig:14}, computed for the representative choice $d/\ell = 0.95$.  The bifurcation diagram for the problem sketched in the $\lambda-\alpha$ plane of equilibrium solutions based on quasi-static calculations accurately identifies the occurrence of the snap-through instability at the fold bifurcation, the critical angle $\alpha_\star$, and the pre- and post-snap configurations labeled ${\rm A}$ and ${\rm C}$, respectively. However, these calculations do not provide information about the dynamic nature of the snap. It is conventional practice to indicate the snap by a path joining ${\rm A}$ and ${\rm C}$ in the equilibrium plane. The path correctly conveys that the snap occurs at the critical angle, but little else is right.  In fact, the dynamic nature of the snap implies that, except for the unstable equilibrium ${\rm A}$ at the start of the snap, no point on the segment ${\rm AC}$ is realized.

The model proposed here augments the bifurcation diagram by computing the snap-dynamics in the $\lambda-\dot{\lambda}$ phase plane positioned at the critical angle. It does so by parameterizing the snap-path by the arch height over the segment ${\rm AC}$. Furthermore, the snap-through trajectory of the center shown in the figure fully defines the dynamic evolution of the arch in the model. The model's prediction ceases at state ${\rm B}$ reached at the end of the snap. Thereafter, the arch vibrates around ${\rm B}$ and may settle to the equilibrium state ${\rm C}$ if its kinetic energy can be dissipated; this regime of the arch's dynamics can be studied using routine modal expansion techniques.

It remains to be seen if the hypothesis underlying the model is also effective in the case of a symmetrically snapping elastic ribbon. For then, it may be possible to generalize the proposed model to predict snap-through profiles of the snapping ribbon, which will help analyze the instability-driven underwater propulsion problem that motivated our work.

\vskip6pt

\enlargethispage{20pt}

\noindent
\paragraph{Acknowledgement:} {We gratefully acknowledge support for
  this work from the Anusandhan National Research Foundation (ANRF)
  through the grant CRG/2020/003641.}

\vskip2pc

\bibliographystyle{RS}
\bibliography{refs}

\end{document}